\newcommand*{\ATLASLATEXPATH}{./}

\documentclass[cernpreprint,texlive=2011,txfonts,UKenglish]{\ATLASLATEXPATH atlasdoc}
 \pdfoutput=1

\usepackage[block=none]{\ATLASLATEXPATH atlaspackage}
\usepackage{multirow}
\usepackage{float}

\usepackage{\ATLASLATEXPATH atlasbiblatex}

\usepackage{\ATLASLATEXPATH atlascontribute}

\usepackage[BSM=true]{\ATLASLATEXPATH atlasphysics}

\graphicspath{{/}{/}}

\usepackage{Xhh4b-defs}

\AtlasTitle{Search for pair production of Higgs bosons in the \fourb final state using proton--proton collisions at $\sqrt{s} = 13$ \TeV{} with the ATLAS detector}

\usepackage{authblk}
\author{The ATLAS Collaboration%
}

\date{\today}

\AtlasRefCode{EXOT-2015-11}

\AtlasNote{ATL-COM-PHYS-2016-104}

\PreprintIdNumber{CERN-EP-2016-142}

\AtlasJournal{PRD}
\AtlasJournalRef{\PRD 94 (2016) 052002}
\AtlasDOI{10.1103/PhysRevD.94.052002}
\AtlasAbstract{%
A search for Higgs-boson pair production in the \fourb\ final state is carried out 
with 3.2~\ifb\ of proton--proton collision data collected at $\rts = 13$ \TeV{} 
with the ATLAS detector.
The data are consistent with the estimated background and are used to set
upper limits on the production cross section of Higgs-boson pairs times branching ratio to \fourb\ 
for both nonresonant and resonant production.
In the case of resonant production of Kaluza--Klein gravitons within
the Randall--Sundrum model, upper limits in the 24 to 91 fb range are 
obtained for masses between 600 and 3000 \GeV{},
at the 95\% confidence level.
The production cross section times branching ratio for nonresonant Higgs-boson pairs 
is also constrained to be less than 1.22 pb, at the 95\% confidence level.
}

\AtlasCoverSupportingNote{Supporting documentation}{https://cds.cern.ch/record/2062404}
\AtlasCoverCommentsDeadline{1 June 2016}

\AtlasCoverAnalysisTeam{
J.~Alison, M.~Bellomo, P.~Bryant, R.~Falla, M.~Franklin, J.~Frost, J.~Guimar\~{a}es da Costa, C.~Issever, M.~Kagan, N.~Konstantinidis, T.~Lazovich, N.~Norjoharuddeen, G.~Rosin, M.~Shochet, B.~Tong, A.~Tuna, D.~Wardrope, S.~Willocq, Q.~Zeng
}

\AtlasCoverEdBoardMember{Markus Elsing~(chair)}
\AtlasCoverEdBoardMember{Tatjana Lenz}
\AtlasCoverEdBoardMember{Chris Pollard}

\AtlasCoverEgroupEditors{atlas-exot-2015-11-editors@cern.ch}

\AtlasCoverEgroupEdBoard{atlas-exot-2015-11-editorial-board@cern.ch}

\hypersetup{pdftitle={ATLAS draft},pdfauthor={The ATLAS Collaboration}}

\begin{document}

\maketitle



\section{Introduction}
\label{sec:introduction}
The discovery of a Higgs boson ($h$) \cite{Aad:2012tfa,Chatrchyan201230} at the 
Large Hadron Collider (LHC) provides
an opportunity to search for physics beyond the Standard Model (SM) in channels
involving couplings with the Higgs boson.
In particular, the production cross section for Higgs-boson pairs in the SM is
significantly smaller than predicted by a host of models, making this channel
promising for a search for new phenomena.
Examples of such models are the bulk Randall--Sundrum (RS) model~\cite{Agashe:2007zd,Fitzpatrick}
with a warped extra dimension
and the two-Higgs-doublet model (2HDM)~\cite{Branco:2011iw}.
In the RS model, spin-2 Kaluza--Klein (KK) excitations of the graviton \Grav are produced
via gluon fusion with subsequent decay to the $hh$ final state.
Similarly, a heavy spin-0 scalar, $H$, in the 2HDM also gives rise to a resonant $hh$ signature.
Enhanced nonresonant $hh$ production is expected in models with light colored scalars
\cite{PhysRevD.86.095023} or direct $t\bar{t}hh$ vertices~\cite{Grober:2010yv,Contino:2012xk}.

Previous searches for $hh$ production have been performed by the ATLAS and CMS collaborations
with $pp$ collisions at $\rts = 8$ \TeV{}.
The final states include \fourb\ ~\cite{Aad:2015uka, Khachatryan:2015yea},
$b\bar{b} \tau^+\tau^-$~\cite{Aad:2015xja, Khachatryan:2015tha},
$b\bar{b}\gamma\gamma$~\cite{Aad:2014yja, Khachatryan:2016sey}, and
$\gamma\gamma W W^\ast$~\cite{Aad:2015xja}.
A combination of these different channels has been performed by ATLAS~\cite{Aad:2015xja},
which shows that 
for resonance masses above 500~\GeV{}
the sensitivity is highest in the \fourb\ channel.

The dominant \hbb\ decay mode is exploited in this paper to search for both resonant
and nonresonant production of Higgs-boson pairs in $pp$ collisions at $\rts = 13$ \TeV{}.
Two analyses are presented. 
The ``resolved'' analysis is optimized for $hh$ systems that have sufficiently
low mass to be resolved into four distinct $b$-jet signatures in the ATLAS detector.
The ``boosted'' analysis focuses on higher-mass $hh$ systems that are characterized
by higher-momentum Higgs bosons for which the two $b$-jets cannot be resolved
due to the high boost. In this situation, large-radius jets are utilized to capture
the by-products of each Higgs-boson decay and small-radius track jets are used to
identify the presence of $b$-hadrons.
The final results are obtained using the resolved analysis
up to resonance masses of 1100 \GeV{}, where its expected sensitivity is higher than
that of the boosted analysis, whereas the boosted analysis is used for
masses above 1100 \GeV{}.

The two analyses generally follow the same approach as that adopted for
the 8~\TeV{} data (see Ref.~\cite{Aad:2015uka}).
The analysis of the 13 \TeV{} data reported in this paper benefits
from an enhanced sensitivity to high-mass resonances due to the significant
increase in the production cross section in that kinematic region.
Furthermore, the boosted analysis includes a channel with only three
$b$-tagged track jets, in addition to the channel with four $b$-tagged track jets
already included in the previous analysis.
This new channel improves sensitivity for resonances with mass above 2000 \GeV{}
because the $b$-jet identification efficiency decreases sharply at high transverse momenta.
The boosted analysis also operates with smaller track-jet radii to account for the larger boost at 13 \TeV{}.

\section{The ATLAS detector}
\label{sec:detector}

\newcommand{\AtlasCoordFootnote}{%
ATLAS uses a right-handed coordinate system with its origin at the nominal interaction point (IP)
in the center of the detector and the $z$-axis along the beam pipe.
The $x$-axis points from the IP to the center of the LHC ring,
and the $y$-axis points upwards.
Cylindrical coordinates $(r,\phi)$ are used in the transverse plane, 
$\phi$ being the azimuthal angle around the $z$-axis.
The pseudorapidity is defined in terms of the polar angle $\theta$ as $\eta = -\ln \tan(\theta/2)$.
Angular distance is measured in units of $\Delta R \equiv \sqrt{(\Delta\eta)^{2} + (\Delta\phi)^{2}}$.}

The ATLAS experiment~\cite{PERF-2007-01} at the LHC is a multipurpose particle detector
with a forward-backward symmetric cylindrical geometry and a near $4\pi$ coverage in 
solid angle.\footnote{\AtlasCoordFootnote}
It consists of an inner tracking detector surrounded by a thin superconducting solenoid
providing a \SI{2}{\tesla} axial magnetic field, electromagnetic and hadronic calorimeters, and a muon spectrometer.
The inner tracking detector (ID) covers the pseudorapidity range $|\eta| < 2.5$.
It consists of silicon pixel, silicon microstrip, and transition radiation tracking (TRT) detectors.
A new innermost pixel layer~\cite{IBL} inserted at a mean radius of 3.3 cm
is used for the first time in the 2015 data taking.
Lead/liquid-argon (LAr) sampling calorimeters provide electromagnetic (EM) energy measurements.
A steel/scintillator-tile hadronic calorimeter covers the central pseudorapidity range ($|\eta| < 1.7$).
The endcap and forward regions are instrumented with LAr calorimeters
for both the EM and hadronic energy measurements up to $|\eta| = 4.9$.
The muon spectrometer (MS) surrounds the calorimeters and includes
three large superconducting air-core toroids.
The field integral of the toroids ranges between \num{2.0} and \SI{6.0}{\tesla\metre} 
for most of the detector.
The MS includes a system of fast detectors for triggering and precision tracking chambers.
A dedicated trigger system is used to select events.
The first-level trigger is implemented in hardware and uses the calorimeter and muon detectors
to reduce the accepted rate to \SI{100}{\kilo\hertz}.
This is followed by a software-based high-level trigger (HLT) that reduces the accepted event 
rate to \SI{1}{\kilo\hertz} on average.

\section{Data and simulation samples}
\label{sec:data-mc}

The data sample used in this analysis was collected during the 2015 LHC run with
$pp$ collisions at $\rts = 13$ \TeV{}.
After requiring that the data be collected during stable beam conditions and
that relevant detector systems be functional, the total integrated luminosity
is estimated to be 3.2 \ifb\ with an uncertainty of 5.0\% derived following
the methodology detailed in Ref.~\cite{Aad:2013ucp}.
In the resolved analysis, events are selected by a combination of three triggers requiring either one or two jets selected by a dedicated HLT $b$-tagging algorithm~\cite{Aad:2015ydr}.
These triggers require either one $b$-tagged jet with transverse momentum $\pt > 225$ \GeV, two $b$-tagged jets with $\pt > 55$ \GeV\ and an additional jet with $\pt > 100$ \GeV, or four jets with $\pt > 35$ \GeV, two of which are $b$-tagged.
A trigger requiring a single jet of radius 1.0 and $\pt > 360$ \GeV{} is used to select events in the boosted analysis.
The \pt\ thresholds for these single- or multiple-jet triggers are lower at the first level of the trigger system.
The combination of all the above triggers has an efficiency rising from 95\% to 99\% for selecting \fourb\ signal
events passing the full analysis selection as the resonance mass increases.

Simulated Monte Carlo (MC) event samples are used to model signal production and
the background from \ttbar\ and $Z$+jets events. A method based on data
is used to model the dominant multijet backgound. Signal \Grav\ events are generated
at leading order (LO) with \MG5amcatnlo v2.2.2~\cite{MG5aMCatNLO} using the NNPDF2.3 LO 
parton distribution function (PDF) set~\cite{Ball:2012cx}, 
and \PYTHIA 8.186~\cite{pythia8} to model the parton shower and hadronization process
using the A14 set of tuned underlying-event parameters~\cite{ATL-PHYS-PUB-2014-021}.
The Higgs-boson mass is set to 125.0 \GeV{}.
Values of the signal cross section times branching ratio for \Gtohhb\ with the
coupling constant $\kMPl = 1$ are 
11.2 fb and 0.185 fb for \Grav\ masses of 1000 \GeV{} and 2000 \GeV{}, respectively. 
The parameter $k$ corresponds to the curvature of the warped extra dimension and
the effective four-dimensional Planck scale $\bar{M}_{\rm{Pl}} = 2.4 \times 10^{18}$ \GeV{}.
Signal samples are also generated with $\kMPl = 2$ to study broader resonances.
Both the cross section and natural width depend on $(\kMPl)^2$.
Generation of the heavy $H$ scalar in a simplified model with a fixed narrow
width $\Gamma_H = 1~\GeV$ is performed with \MG5amcatnlo
and the CT10 PDF set~\cite{Lai:2010vv}. 
With this $\Gamma_H$ choice, the width of the reconstructed $hh$ resonance is dominated by the
experimental resolution.
For this model, parton showering and hadronization are handled
by \herwigpp~\cite{Bahr:2008pv} with the CTEQ6L1 PDF set~\cite{cteq6l1} and the UEEE5
underlying-event tune~\cite{Gieseke:2012ft}.
The scalar interpretation for this search only makes use of the acceptance times
efficiency from this model and no interpretation in terms of 2HDM
parameters is presented.
Nonresonant SM \pptofourb\ events are generated via the gluon-fusion
process with \MG5amcatnlo\ using form factors for the top-quark loop from 
HPAIR~\cite{PhysRevD.58.115012, Plehn199646}.
The cross section times branching ratio to the \fourb\ final state, 
evaluated at next-to-next-to-leading order with the summation of logarithms 
at next-to-next-leading-logarithm accuracy, is $11.3^{+0.9}_{-1.0}$~fb~\cite{LHCHXSWG}.
The uncertainty includes the effects due to renormalization and factorization scales,
PDF set, $\alpha_{\mathrm{S}}$, effects of finite top-quark mass in loops, and the \hbb\ branching 
ratio.

Generation of \ttbar\ events is performed with \powhegbox\ v1 using the
CT10 PDF set.
The parton shower, hadronization, and the underlying event
are simulated using \PYTHIA 6.428~\cite{Sjostrand:2006za} with the CTEQ6L1 PDF 
set and the corresponding Perugia 2012 tune~\cite{Skands:2010ak}. 
The top-quark mass is set to 172.5~\GeV{}.
Higher-order corrections to \ttbar\ cross sections are computed with 
Top++ 2.0~\cite{Czakon:2011xx}. These incorporate NNLO corrections in QCD, including
resummation of next-to-next-to-leading logarithmic soft gluon terms.
The overall \ttbar\ normalization is extracted from the data while the shape
of kinematic distributions is taken from MC simulation.
The $Z$+jets sample is generated using \PYTHIA 8.186 with the 
NNPDF2.3 LO PDF set.

For all MC samples,
charm-hadron and bottom-hadron decays are handled by {\textsc{EvtGen}} 1.2.0~\cite{EvtGen}.
To simulate the impact of multiple $pp$ interactions that occur within the same
or nearby bunch crossings (pileup), minimum-bias events generated with \PYTHIA~8
are overlaid on top of the hard scatter event.
The detector response is simulated with \GEANT~4~\cite{Agostinelli:2002hh, Aad:2010ah}
and the events are processed with the same reconstruction software as that used for
the data.

\section{Event reconstruction}
\label{sec:reconstruction}
The resolved and boosted analyses rely on the reconstruction of jets with the
anti-\kt clustering algorithm~\cite{Cacciari:2008gp} but with different values of 
the radius parameter $R$.
Calorimeter jets with $R = 0.4$ ($1.0$) 
are used to determine the kinematic properties of Higgs-boson candidates
in the resolved (boosted) analysis.
Those jets are reconstructed from topological clusters of energy 
deposits in calorimeter cells.
The $R = 0.4$ jet energies are determined from reconstructed cluster energies
at the electromagnetic scale with correction factors derived 
from simulation to account for the response of the calorimeter to hadrons~\cite{Aad:2011he}.
Jets from pileup are suppressed with the use of tracking information as detailed in Ref.~\cite{JVT}.
The $R = 1.0$ jets are built from locally calibrated clusters~\cite{Aad:2011he} and
are trimmed~\cite{Krohn2010} to minimize the impact of pileup.
This trimming proceeds by reclustering the jet with the 
\kt algorithm~\cite{Ellis:1993tq} into smaller 
$R = 0.2$ subjets and removing those subjets with $\pt^{\mathrm{subjet}}/\pt^{\mathrm{jet}} < 0.05$, where $\pt^{\mathrm{subjet}}$ is the transverse momentum of the subjet and 
$\pt^{\mathrm{jet}}$ that of the original jet.
In addition to the above large-$R$ trimmed jets, the boosted analysis uses track jets with 
$R = 0.2$ to identify $b$-hadrons from Higgs-boson decays~\cite{ATL-PHYS-PUB-2015-035}. 
Such jets are reconstructed from charged-particle tracks  
with $\pt > 0.4$ \GeV{} and $|\eta| < 2.5$
that satisfy a set of hit and impact parameter criteria to make sure that the tracks 
originate from the primary vertex, thereby minimizing the impact of pileup.
Track jets are associated to large-$R$ jets using ghost association~\cite{Cacciari:2007fd}.
In this method, the large-$R$ jet algorithm is rerun with both the four-momenta of track
jets modified to have infinitesimally small momentum (the ``ghosts'') and 
all topological energy clusters in the event as potential constituents of jets.
As a result, the presence of track jets does not alter the large-$R$ jets already found and
their association to specific large-$R$ jets is determined by the jet algorithm. 
Collision vertices are reconstructed requiring a minimum of two tracks with
$\pt > 0.4$ \GeV{} in each vertex. 
The primary vertex is chosen to be the vertex with the largest $\sum \pt^2$,
where the sum extends over all tracks associated with the vertex.

The identification of jets containing $b$-hadrons is based on the $R = 0.4$ calorimeter 
($R = 0.2$ track) jets in the resolved (boosted) analysis and a multivariate tagging 
algorithm~\cite{ATL-PHYS-PUB-2015-022}.
This algorithm is applied to a set of tracks with loose impact parameter constraints
in a region of interest around each jet axis to enable the reconstruction of the
$b$-hadron decay vertex.
The $b$-tagging requirements result in an efficiency of 70\% (77\%) for jets containing
$b$-hadrons in the resolved (boosted) analysis, as determined in a sample
of simulated \ttbar\ events. 
The corresponding efficiencies for $c$-hadron jets and light-quark or gluon jets
are 12\% (29\%) and 0.2\% (1.4\%), respectively.
Different $b$-tagging operating points are chosen
in the two analyses to maximize their respective sensitivities.

Muons are reconstructed by combining tracks in the ID and
MS, and are required to satisfy tight muon identification
criteria~\cite{Aad:2016jkr}.
The four-momentum of muons with $\pt > 4$~\GeV{} and $|\eta| < 2.5$, that are within 
\DR\ of 0.4 (0.2) of jets used for $b$-tagging in the resolved (boosted) analysis, 
is added to the calorimeter jet four-momentum to partially account 
for the energy lost in semileptonic $b$-hadron decays.

\section{Event selection}
\label{sec:selection}
The event selection for the resolved and boosted analyses is described below.
These analyses are optimized independently for the reconstruction
and selection of $hh \to b\bar{b} b\bar{b}$ final states,
with the resolved analysis aiming at event topologies containing four
distinct $b$-jets, whereas the boosted analysis focuses on topologies
with higher-momentum Higgs bosons resulting in merged jets.

Different selection and background estimation strategies are adopted for
the two analyses. To facilitate the comparison between these different choices,
Table~\ref{tab:analysis} summarizes each of the requirements and approaches
described in this section.

\begin{table}[t]
\begin{center}
\begin{tabular}{l  l  l }
\toprule
                       &   Resolved                                                   &    Boosted   \\
\midrule
Event preselection     &   $\geq 4$ jets  with                                        &  $\geq 2$ large-$R$ jets  with \\
                       &   ~~~$\pt > 40$ \GeV{},                                      &  ~~~$350\,(250) <\pt < 1500$ \GeV{}, \\ 
                       &   ~~~$|\eta|<2.5$                                            &  ~~~$|\eta|<2.0$, $m_{\mathrm{J}}>50$ \GeV{} \\ 
                       &   $\geq 2$ dijets with                                       &  $\geq 2$ track jets associated to  \\
                       &   ~~~$\pt > 200\,(150)$ \GeV{}, $\Delta R < 1.5$,            & ~~~each large-$R$ jet with \\
                       &   ~~~$\pt > f(\mfourj)$, $|\Delta\eta| < f'(\mfourj)$        & ~~~$\pt>10$ \GeV{}, $|\eta|<2.5$, $|\Delta\eta|<1.7$ \\
\midrule
\ttbar veto               &   $X_{tt} < 3.2$                                              &   -- \\
\midrule
Tagging                &   $4$ $b$-tagged jets                                        &  $3$ or $4$ $b$-tagged jets \\
\midrule
Signal region (SR)     &   \multicolumn{2}{c}{$X_{hh} < 1.6$} \\
Sideband region (SB)   &   \multicolumn{2}{l}{Resolved: $\sqrt{(\mlead - 124\,\GeV)^2 + (\msublead - 115\,\GeV)^2} > 58\,\GeV$}  \\
                       &   \multicolumn{2}{l}{Boosted: ~~$\sqrt{(\mleadJ - 124\,\GeV)^2 + (\msublJ - 115\,\GeV)^2} > 36\,\GeV$}  \\
Control region (CR)    &   \multicolumn{2}{c}{complementary to SR and SB} \\
\midrule
Multijet normalization &   scaled yields from 2-tag SR, & scaled yields from 2-tag SR, \\
                       &    scaling derived from    &  scaling derived from \\
                       &   4-tag to 2-tag ratio in SB                             &  3(4)-tag to 2-tag fit \\
                       &                                                          &  to leading jet mass in SB  \\
\midrule
Multijet shape         &   \multicolumn{2}{c}{derived from 2-tag SR}   \\
\midrule
$t\bar{t}$ normalization & scaled yields from $t\bar{t}$ CR,                      & scaled yields from MC simulation,  \\
                         & scaling derived from                                   & scaling derived from \\
                         & semileptonic $t\bar{t}$ events                         & 3(4)-tag to 2-tag fit to leading \\
                         &                                                        & jet mass in SB  \\                                  
\midrule
$t\bar{t}$ shape       &   \multicolumn{2}{c}{derived from MC simulation} \\
\bottomrule
\end{tabular}
\caption{Event selection requirements and definition of the different regions used in the resolved and boosted analyses.
The methodologies used to estimate the background normalization and shape are also outlined. The variables are defined in the text. Dijet and large-$R$ jet
minimum \pt\ values are indicated for leading (subleading) such objects.
The functions $f(\mfourj)$ and $f'(\mfourj)$ represent the mass dependence of 
the minimum \pt\ and maximum $|\Delta\eta|$ requirements placed on the dijet 
candidates in the resolved analysis.
}
\label{tab:analysis}
\end{center}
\end{table}

\subsection{Resolved analysis}
\label{sec:resolved}

\subsubsection{Selection}
\label{sec:ResolvedSel}

Events selected for the resolved analysis must contain at least four $b$-tagged jets with $|\eta| < 2.5$ and  $\pt~>~40$ \GeV.
The four highest-\pt~$b$-tagged jets are used to form two dijet systems, requiring an angular distance $\Delta R$ between the jets within the dijet system smaller than 1.5. 
The transverse momentum of the leading (subleading) dijet system is required to be greater than 200 (150)~\GeV.
These requirements are made to ensure a high trigger efficiency and to avoid ambiguities in forming dijets.
In the rare case that a jet is assigned to more than one dijet system, only the combination containing the jets with the highest probability of being $b$-jets according to the multivariate $b$-tagging algorithm is considered.

The resolved analysis considers resonance masses in the range 400--1500 \GeV. 
Event selection that varies as a function of the reconstructed resonance mass (\mfourj) is used to increase the analysis sensitivity across the mass range searched.
Mass-dependent selection requirements are made on the leading dijet \pt, the subleading dijet \pt and the pseudorapidity difference between the dijets as follows~\cite{Aad:2015uka}:

\[\pt^{\mathrm lead} > \begin{cases}
		400\,\GeV& \mathrm{if}\ $\mfourj$ > 910\,\GeV, \\
		200\,\GeV& \mathrm{if}\ $\mfourj$ < 600\,\GeV, \\ 
		0.65\,\mfourj - 190\,\GeV& \mathrm{otherwise},
		\end{cases}
\]
\[\pt^{\mathrm subl} > \begin{cases}
	260\,\GeV& \mathrm{if }\ $\mfourj$ > 990\,\GeV, \\
	150\,\GeV& \mathrm{if }\ $\mfourj$ < 520\,\GeV, \\ 
	0.23\,\mfourj + 30\,\GeV& \mathrm{otherwise},
	\end{cases}
\]
\[ \dEta < \begin{cases}
	1.0 & \mathrm{if }\ $\mfourj$ < 820\,\GeV, \\
	1.6\times10^{-3} \mfourj - 0.28 & \mathrm{otherwise}.
	\end{cases}
\]

These selection requirements were optimized simultaneously by performing a three-dimensional scan of threshold values, using the expected exclusion limit on the \Grav resonance with $\kMPl=1$ as a metric.

After selecting two dijets that satisfy the mass-dependent criteria, 15\% of the total background consists of \ttbar events.
This \ttbar\ background mainly comprises events where both top quarks decay hadronically. 
These hadronic decays often lead to three jets for each top quark -- one $b$-jet directly from the top-quark decay and two from the decay of the $W$ boson. 
Reduction of the \ttbar background is important as relatively large systematic uncertainties are associated with modeling \ttbar in the signal region.
In order to reduce the \ttbar\ background, jets not already used in the formation of the two dijets (``extra jets'') in the event are used to reconstruct $W$-boson and top-quark candidates by combining them with one or both of the jets in a given dijet. 
These extra jets are required to have $\pt > 30$~\GeV, $|\eta| < 2.5$, and $\Delta R<1.5$ relative to the dijet. 
The $W$-boson candidate is reconstructed by adding the four-momentum of the extra jet to the four-momentum of the jet in the dijet system with the lowest probability of being a $b$-jet according to the multivariate $b$-tagging algorithm. 
The top-quark candidate is reconstructed by summing the dijet and the extra jet. 
The compatibility with the top-quark decay hypothesis is then determined using the variable:

\begin{equation}
X_{tt}\,=\,\sqrt{\left (\frac{m_{W}\,-\,80.4\,\GeV}{0.1\,m_{W}}\right )^2 + \left (\frac{m_{t}\,-\,172.5\,\GeV}{0.1\,m_{t}}\right )^2}~,
\end{equation}

\noindent
where $m_{W}$ and $m_t$ are the invariant masses of the $W$-boson and top-quark candidates.
The values in the denominator approximate the dijet and three-jet system mass resolutions. 
If either dijet in an event has $X_{tt} < 3.2$ for any possible combination with an extra jet, the event is rejected. 
This requirement, referred to as the ``\ttbar\ veto'', reduces the \ttbar\ background by $\sim$60\%, while retaining $\sim$90\% of signal events.
The event selection criteria described above are collectively referred to as the ``4-tag'' selection requirements. 

Following the 4-tag selection, a requirement on the combination of the leading and subleading dijet masses (\mlead and \msublead, respectively) is used to define the signal region. 
The signal region is defined using the variable:

\begin{equation}
X_{hh}\,=\,\sqrt{\left (\frac{\mlead\,-\,124\,\GeV}{0.1\,\mlead}\right )^2 +\, \left (\frac{\msublead\,-\,115\,\GeV}{0.1\,\msublead}\right )^2}~, 
\label{eq:Xhh}
\end{equation}
\noindent

where the $0.1\,m_{\mathrm 2j}$ terms approximate the widths of the mass distributions. 
The center of the signal region was optimized using \Grav\ samples with $\kMPl=1$. 
On average, the subleading Higgs-boson candidate is reconstructed at lower masses as a result of energy lost from semileptonic $b$-hadron decays and final-state radiation. 
The signal region is defined as $X_{hh} < 1.6$. 
This corresponds to the kinematic requirements illustrated by the inner region in Figure~\ref{fig:resolved_m12m34}.
The data shown in this figure are derived from a sample of events that satisfy all selection criteria except for having only two jets that pass the $b$-tagging requirements, referred to as the ``2-tag'' sample.

\begin{figure}[!ht]
\begin{center}
\includegraphics[width=0.60\textwidth]{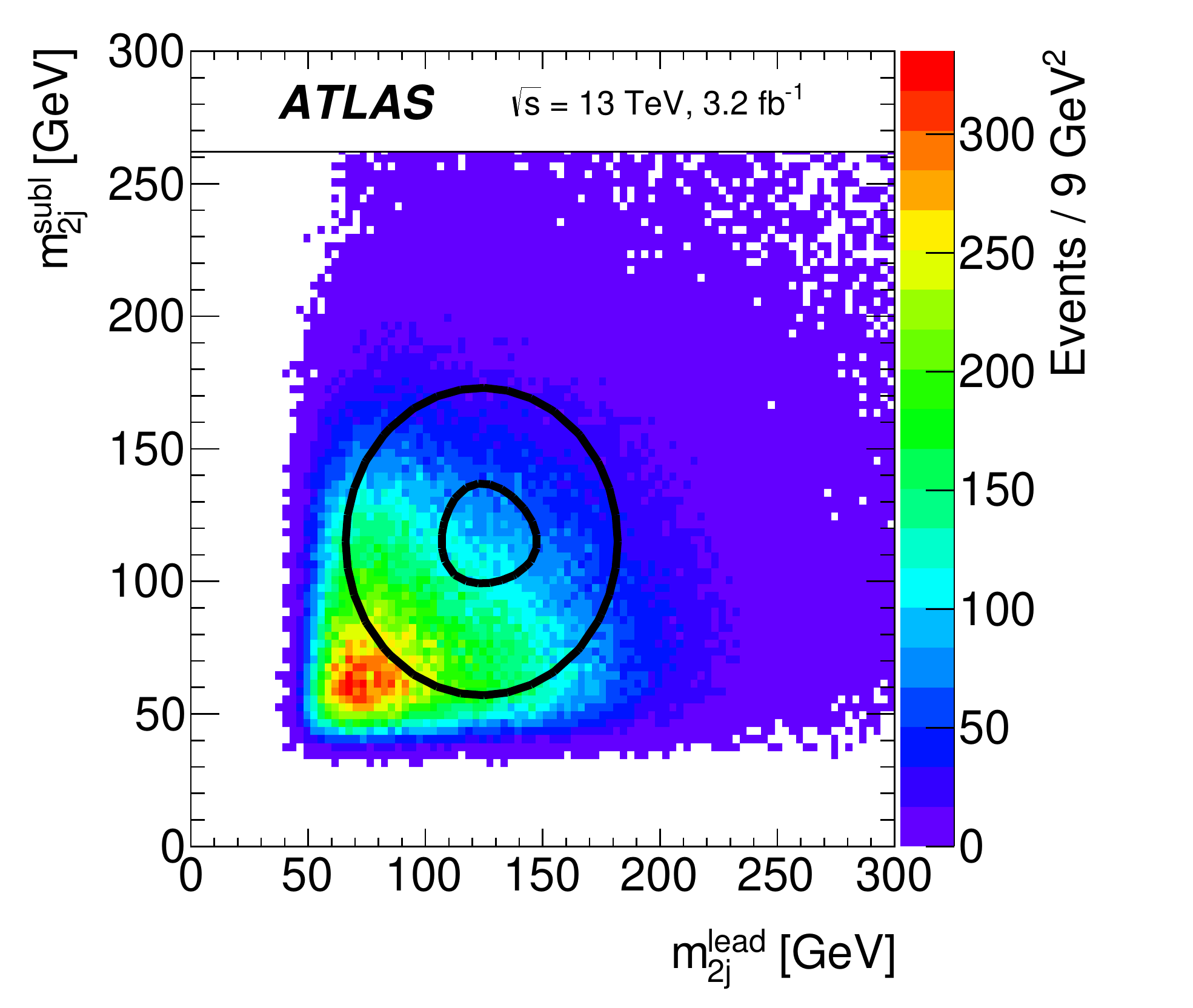}
\caption{The \msublead vs. \mlead distribution for the 2-tag data sample used to model the multijet background in the resolved analysis. 
        The signal region is the area surrounded by the inner black contour line, centered on \mlead=124\,\GeV, \msublead=115\,\GeV. 
        The control region is the area inside the outer black contour line, excluding the signal region. 
        The sideband region is the area outside the outer contour line.}
\label{fig:resolved_m12m34}
\end{center}
\end{figure}

The acceptance times efficiency for each stage of the resolved-analysis event selection is shown in Figure \ref{fig:cutflowPlotsResolved} for spin-2 and spin-0 resonances.
The acceptance times efficiency, \accEff, of the full selection for the \Grav with $\kMPl = 1$ ranges from 0.1\% for a \Grav of mass 400 \GeV\ to 5.3\% for a \Grav with a mass of 1000 \GeV.
The spin of the resonance affects the angular distribution of the decay products, resulting in a lower acceptance in the case of a spin-0 $H$ boson than for the spin-2 \Grav.
As a result, the spin-0 resonance search is performed starting at a mass of 500 \GeV.
Nonresonant di-Higgs production occurs primarily at the low end of the \mfourj spectrum, leading to $\accEff = 0.64\%$ for the full selection.

\begin{figure}[!ht]
\begin{center}
\includegraphics[width=0.49\textwidth]{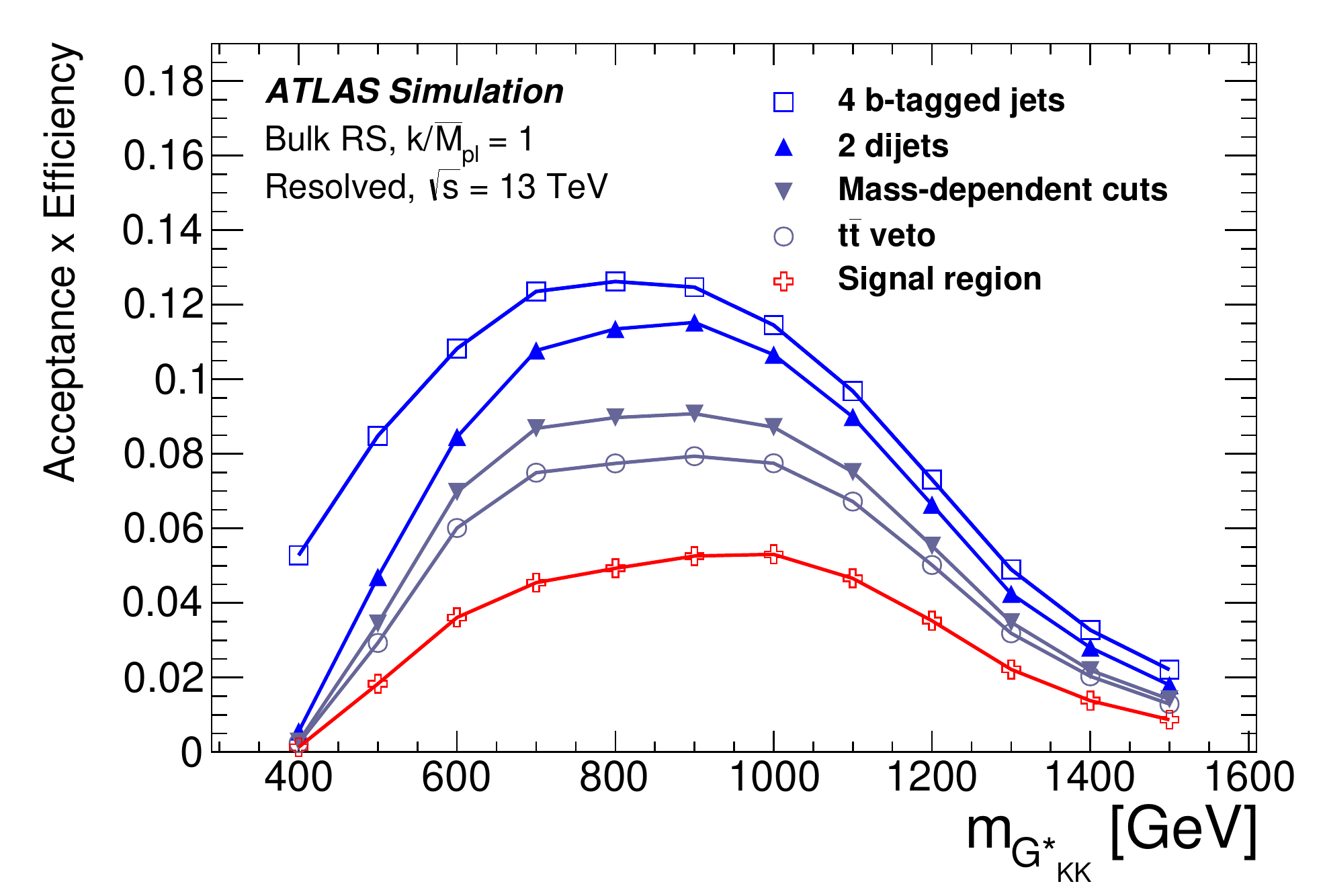}
\includegraphics[width=0.49\textwidth]{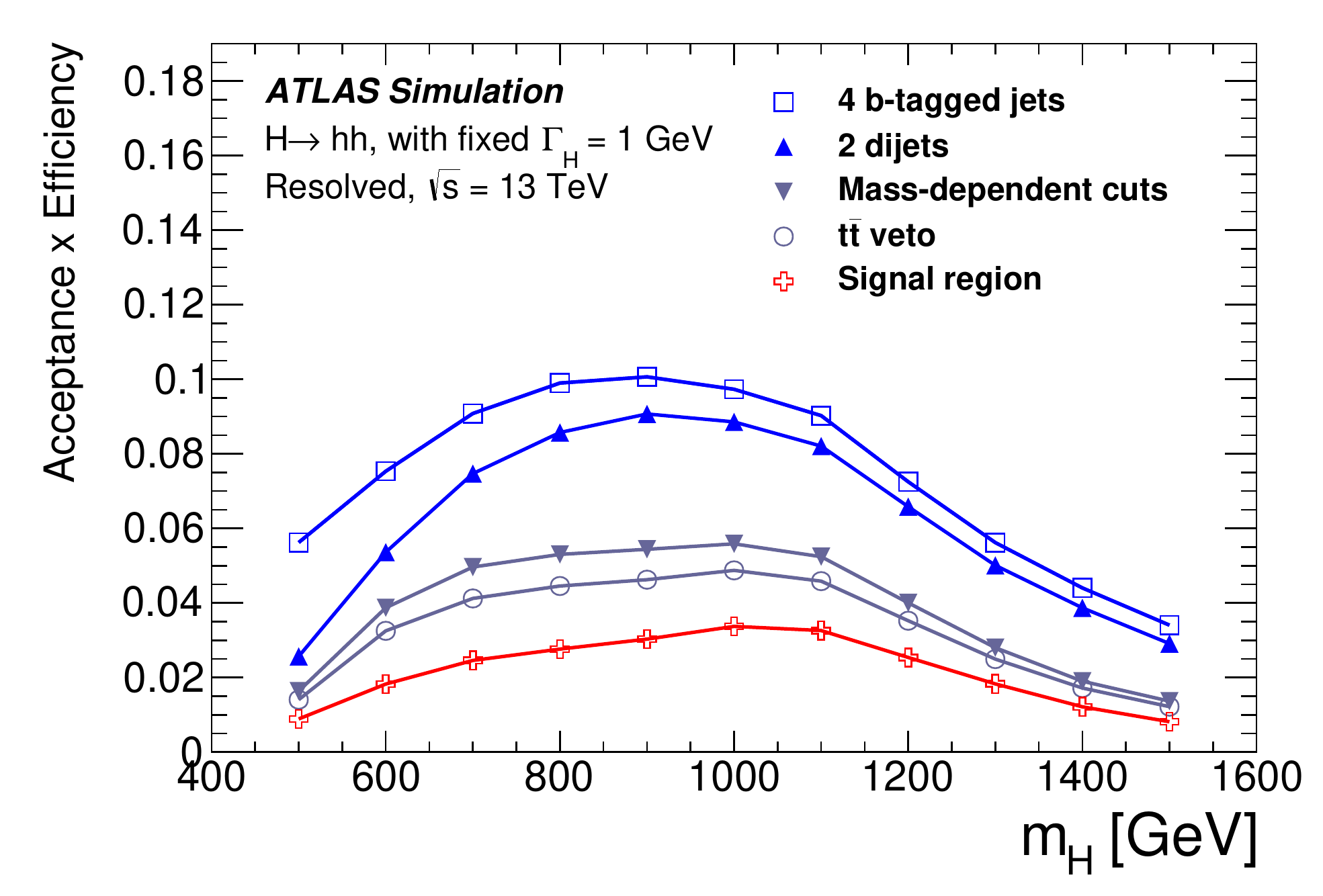}
\caption{The selection efficiency as a function of resonance mass at each stage of the event selection for (left) \Gtohhb and (right) \Htohhb decays in the resolved analysis.}
\label{fig:cutflowPlotsResolved}
\end{center}
\end{figure}

The final step in the resonant analysis is to search for an excess in the \mfourj 
distribution for events in the signal region. 
The sensitivity of the search is increased by improving the \mfourj resolution in this region. This is achieved by scaling the four-momentum of each of the Higgs-boson candidates
such that their mass is equal to the Higgs-boson mass.
This leads to an improvement of $\sim30\%$ in the signal \mfourj resolution with little impact on the background.

\subsubsection{Background estimation}
\label{sec:ResolvedBkg}
After the 4-tag selection described above, $\sim$90\% of the remaining background in the signal region originates from multijet events, which are modeled using data. 
The remaining $\sim$10\% of the background is expected from \ttbar\ events. 
The \ttbar\ yield is determined from data, while the \mfourj~shape is taken from MC simulation. 
The $Z$+jets contribution is less than 1\% of the total background and is estimated from MC simulation.
The background from all other sources -- including processes featuring Higgs bosons -- is negligible.

\subsubsection*{Multijet background}
The multijet background is modeled using an independent data sample selected using the same trigger and selection requirements as described above, 
except for the $b$-tagging requirement: only one of the two selected dijets is formed from $b$-tagged jets, while the other dijet is formed from jets that both fail the $b$-tagging requirements. 
This ``2-tag'' selection yields a data sample that consists of 98\% multijet events and 2\% \ttbar\ events. 
The predicted signal contamination is negligible.

The 2-tag sample is normalized to the 4-tag sample and its kinematics corrected for differences introduced by the additional $b$-tagging requirement on the 4-tag sample. 
These kinematic differences arise because the $b$-tagging efficiency varies as a function of jet \pt and $\eta$, the various multijet processes contribute in different fractions, and the fraction of events passed by each trigger path changes. 
The normalization and kinematic corrections are determined using a signal-free sideband region of the \mlead--\msublead plane.
The resulting background model is verified and the associated uncertainties are estimated using a control region. 
The sideband and control regions are shown in Figure \ref{fig:resolved_m12m34}. 
The sideband region is defined as $\sqrt{(\mlead - 124\,\GeV)^2 + (\msublead - 115\,\GeV)^2} > 58\,\GeV$, while the control region is defined as the region in the \mlead--\msublead plane between the signal and sideband regions. 
These definitions are chosen to be orthogonal to the signal region and to give approximately equal event yields in both the sideband and control regions. 

The normalization of the multijet background prediction is set by scaling the number of events in each
region of the 2-tag sample by the following factor \muqcd calculated in the sideband region:

\begin{equation}
\muqcd = \frac{N^{\text{4-tag}}_{\text{Multijet}}}{N^{\text{2-tag}}_{\text{Multijet}}} = \frac{N^{\text{4-tag}}_{\text{data}} - N^{\text{4-tag}}_{t\bar{t}} - N^{\text{4-tag}}_{Z+\text{jets}}}{N^{\text{2-tag}}_{\text{data}}-  N^{\text{2-tag}}_{t\bar{t}} - N^{\text{2-tag}}_{Z+\text{jets}}  },
\label{eqn:qcdnorm}
\end{equation}

\noindent
where $N_{\text{data}}^{\text{2-/4-tag}}$ is the number of events observed in the sideband region in the 2- or 4-tag data sample, respectively. 
The yields $N_{t\bar{t}}^{\text{2-/4-tag}}$ are the estimated number of \ttbar\ events in the 2-/4-tag selected sideband region estimated from MC simulation. 
To predict the distributions of the multijet background in each region, the predicted \ttbar\ 2-tag distributions are first subtracted from the 2-tag data sample before the distribution is scaled by \muqcd. 

The correction for the kinematic differences between 2-tag and 4-tag samples is performed by reweighting events in the 2-tag sample.
The weights are derived in the sideband region, from linear fits to the ratio of the total background model to data for three kinematic distributions that are found to have the largest disagreement between 2-tag and 4-tag: the leading dijet \pt, the angular separation between the jets in the subleading dijet, and the angular separation between the two dijets. 
The reweighting is performed using one-dimensional distributions but is iterated so that correlations between the three variables are taken into account. 
After the correction process, there is agreement between the background model and sideband region data. 

The multijet background model is validated in the control region.
Table~\ref{tab:resolvedSBCR} compares the observed data yield in the control region with the corresponding background estimate.
The modeling of the \mfourj distribution in the control region is shown in Figure~\ref{fig:ctrlregfit}.
The 4-tag events in the control region are well-described by the background model in both normalization and \mfourj shape.

\begin{table}
\begin{center}
\begin{tabular}{ l c  c  }
\toprule
 Sample & Sideband Region & Control Region  \\ 
\midrule
Multijet      &    485.1 $\pm$ 2.1    &     401.5   $\pm$ 2.0 \\
\ttbar        &  ~~~~9.6 $\pm$ 0.9    &    ~~14.0   $\pm$ 1.2 \\
$Z$+jets      &  ~~~~3.1 $\pm$ 0.7    &   ~~~~4.9   $\pm$ 1.0 \\
\midrule                                  
Total         &    497.8 $\pm$ 2.4    &     420.3   $\pm$ 2.5 \\
\midrule                                  
Data          &    496              &     396  \\
\bottomrule
\end{tabular}
\caption{The number of events in data and predicted background events after applying the \ttbar\ veto in the sideband and control regions for the resolved analysis. 
  The uncertainties are purely statistical. 
  The \ttbar\ yield in this table, in contrast to the final result, is estimated using MC simulation.}
\label{tab:resolvedSBCR} 
\end{center}
\end{table}

\begin{figure}[!ht]
\begin{center}
\includegraphics[width=0.58\textwidth]{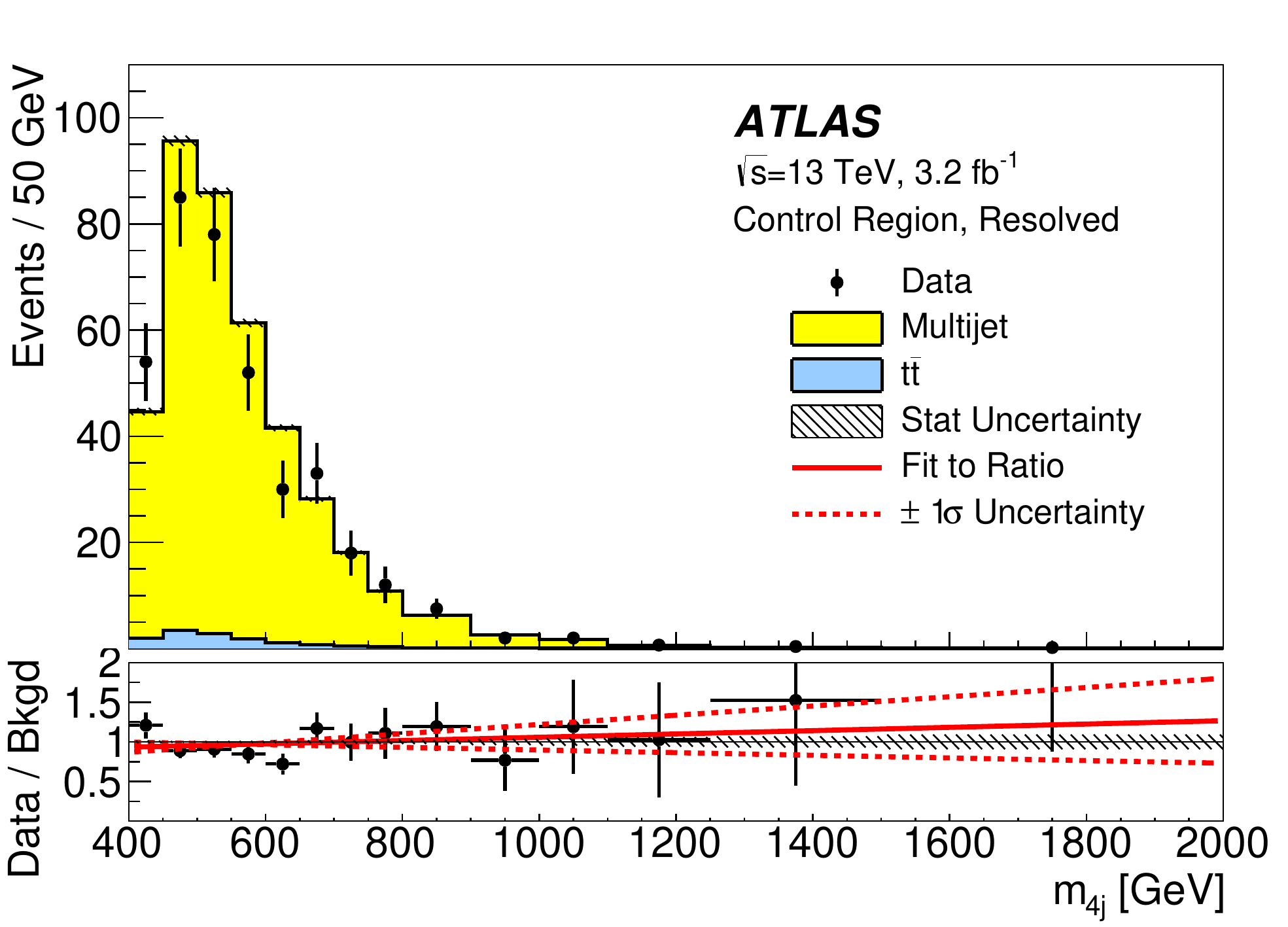}
\caption{The \mfourj distribution in the control region of the resolved analysis for the data and the predicted background (top panel). 
         The small hatched bands drawn on the histogram and on the horizontal line in
the data to background ratio (bottom panel) represent the statistical uncertainty in the total background estimate. 
         The bottom panel also includes a first-order polynomial fit to the data-to-background ratio. 
         The dashed lines show the $\pm 1\sigma$ uncertainties in the two fitted parameters.}
\label{fig:ctrlregfit}
\end{center}
\end{figure}

\subsubsection*{\ttbar background}
\label{sec:ttbarBack}
The normalization of the \ttbar\ background is derived from data in a \ttbar\ control region.
Due to the limited yield in this control region, the shape of the \ttbar\ background is taken from MC simulation.
To further decrease statistical uncertainties, the \ttbar\ shape is derived from MC simulation using the ``2-tag'' selection, with a systematic uncertainty assigned to cover the differences between the 2-tag and 4-tag \mfourj distributions.

The \ttbar\ control region is formed from events which pass the 4-tag selection, except for the \ttbar\ veto, which is reversed: if either of the dijets fails the $X_{tt}$ requirement, the event enters the \ttbar control region.
This selection leads to a sample of 21 events, of which 13.3 are estimated to be multijet events using the 2-tag sample described previously.
After subtracting the multijet background, the \ttbar\ control region yield is extrapolated to predict the \ttbar\ yield in the signal region, $N_{t\bar{t}}$, using the following equation: 
 
\begin{equation}
  N_{t\bar{t}} = \frac{\epsilon_{t}^{2}}{1-\epsilon_{t}^{2}} \times N_{t\bar{t}}^{\text{CR}},
\label{eq:ttbarPred}
\end{equation}
\noindent
where $N_{t\bar{t}}^{\text{CR}}$ is the number of events in the \ttbar\ control region, after subtraction of multijet background, and $\epsilon_{t}$ is the efficiency for a selected dijet in a \ttbar\ event to pass the \ttbar\ veto. 
This equation relies on the assumption that the $\epsilon_{t}$ of each dijet in the event is uncorrelated, an assumption validated in \ttbar\ MC simulation. 
The $\epsilon_{t}$ is measured using an independent, high-purity ``semileptonic \ttbar\hspace{0.02em}'' data sample.
Events in this sample are selected by requiring one dijet candidate passing the nominal selection with $\pt > 150$\,\GeV\ and one ``leptonic top-quark'' candidate. 
The leptonic top-quark candidate is defined using a reconstructed muon and one $b$-tagged jet.
This $b$-tagged jet is required to be distinct from jets in the dijet candidate, and the muon is required to have \pt~$>$ 25\,\GeV, be isolated, and fall a distance $\Delta R < 1.5$ of the $b$-tagged jet. 
The leptonic top-quark candidate is required to have \pt~$>$ 150\,\GeV, where the leptonic top $\vec{p}_{\mathrm{T}}$ is defined as the vector sum of the $b$-jet $\vec{p}_{\mathrm{T}}$ and the muon $\vec{p}_{\mathrm{T}}$.
The \ttbar\ veto efficiency is then measured as the fraction of the reconstructed dijet candidates which passed the \ttbar\ veto, yielding $\epsilon_{t\bar{t}} = $ 0.60 $\pm$ 0.04 (stat.) $\pm$ 0.06 (syst.). 
A 10\% systematic uncertainty is assigned to cover potential differences between $\epsilon_{t}$ as measured in the semileptonic \ttbar\ sample and $\epsilon_{t}$ in the full 4-tag selection, where the method is applied in \ttbar\ MC simulation to evaluate such differences.
The measured $\epsilon_{t}$ agrees well with the corresponding semileptonic \ttbar\ MC prediction of $0.58$.

Equation~(\ref{eq:ttbarPred}) gives a data-driven \ttbar\ background prediction of $4.2\pm 3.8$ events.
The uncertainty is dominated by the statistical uncertainty in the yield in the \ttbar control region, with a smaller contribution from the uncertainty in the measured \ttbar\ veto efficiency.

\subsubsection{Systematic uncertainties}
\label{sec:ResolvedSyst}
Two classes of systematic uncertainties are evaluated: those affecting the modeling of the signal and those affecting the background prediction. 

The signal modeling uncertainties comprise: theoretical uncertainties in the acceptance, uncertainties in the jet energy scale (JES) and resolution (JER), and uncertainties in the $b$-tagging efficiency. 

The following sources of theoretical uncertainty in the acceptance are evaluated: missing higher-order terms in the matrix elements and PDF set, as well as modeling of the underlying event, hadronic showers, initial- and final-state radiation. The total theoretical uncertainty is dominated by the uncertainties associated with the modeling of the initial- and final-state radiation.

The jet energy uncertainties are derived based on in situ measurements performed during Run 1 and from MC simulation extrapolations from Run-1 to Run-2 conditions~\cite{ATL-PHYS-PUB-2015-015}.
The JES systematic uncertainty is evaluated using three separate and orthogonal uncertainty components~\cite{Aad:2014bia}. 
The JER uncertainty is evaluated by smearing jet energies according to the systematic uncertainties of the resolution measurement~\cite{Aad:2014bia}. 
The uncertainty in the $b$-tagging efficiency is evaluated by  propagating the systematic uncertainty in the measured tagging efficiency for $b$-jets~\cite{ATLAS-CONF-2014-004}.
The efficiencies are measured as a function of $b$-jet \pt and \eta.
For $b$-jets with \pt~$> 300$\,\GeV, systematic uncertainties in the tagging efficiencies are extrapolated with MC simulation and are consequently larger~\cite{Aad:2015ydr}.

Systematic uncertainties in the normalization and shape of the multijet background model are assessed in the control region. 
The background prediction in the control region agrees with the observed data to within $\pm5\%$, which is taken as the uncertainty in the predicted multijet yield. 
To further test the robustness of the background estimation, the background model is re-evaluated using different sideband and control region definitions and different $b$-tagging requirements on the ``2-tag'' sample. 
These changes affect the kinematic and flavor compositions of the various regions used in the background prediction.
The control region agreement and signal region predictions of all variations considered are all consistent to within the assigned $\pm5\%$ uncertainty in the multijet background prediction.

The uncertainty in the description of the multijet \mfourj distribution is determined by comparing the background prediction to the data in the control region as shown in Figure~\ref{fig:ctrlregfit}.
To evaluate the level of agreement, a linear fit is performed on the ratio of the distributions. 
This fit, along with its uncertainties, shown in the bottom panel of Figure~\ref{fig:ctrlregfit}, gives a slope consistent with zero. 
The uncertainty in the multijet background shape is defined using the uncertainty in the fitted slope.

The uncertainty in the \ttbar\ normalization is described above.
The uncertainty in the MC-derived \ttbar\ \mfourj distribution is dominated by the uncertainty associated with using the 2-tag selection to model the 4-tag selection. 
This uncertainty is assessed by comparing the 2-tag and 4-tag \ttbar\ MC predictions in the signal region.

Table~\ref{tab:resolvedSyst} summarizes the relative impact of the uncertainties in the event yields. 
 
\begin{table}[ht!]
\begin{center} 
\begin{tabular}{  l  c  c c c  c c }
\toprule
 Source          & Background   & SM $hh$   & \Grav(500 \GeV)    &  \multicolumn{2}{c}{\Grav(800 \GeV)}        & $H$ (800 \GeV)  \\ 
                 &              &           & \frackMPl = 1      & \frackMPl = 1 &  \frackMPl = 2 &       \\ 
\midrule                                      
   Luminosity    &       --     & ~~$5$     &    ~~$5$           &    ~~$5$      & ~~$5$          & ~~$5$  \\
   JER           &       --     & ~~$2$     &    ~~$3$           &    ~~$3$      & ~~$3$          & ~~$4$  \\
   JES           &       --     & $12$      &      $14$          &    ~~$5$      & ~~$4$          & ~~$6$  \\
   $b$-tagging   &       --     & $18$      &      $15$          &      $26$     & $27$         & $26$  \\
   Theoretical   &       --     & ~~$9$     &    ~~$2$           &    ~~$3$      & ~~$3$          & ~~$3$  \\      
   Multijet      &      $5$     & --        &      --            &      --       &       --     & --     \\
   \ttbar        &      $6$     & --        &      --            &      --       &       --     & --     \\
   \midrule                                                      
   Total         &      $8$     & $24$      &      $21$          &     $28$      & $28$         & $28$ \\
\bottomrule
\end{tabular}
\caption{Summary of systematic uncertainties (expressed in percentage yield) in the total background and signal event yields in the signal region of the resolved analysis.
 Uncertainties are provided for nonresonant SM Higgs pair production, for a \Grav 
 resonance with \kMPl = 1 and $m = 500$ \GeV, and for three resonances with $m = 800$ GeV: a \Grav resonance with \kMPl = 1, a \Grav resonance with \kMPl = 2, and a spin-0 narrow-width $H$ boson. }
\label{tab:resolvedSyst}
\end{center}
\end{table}

\subsubsection{Event yields}
\label{sec:ResolvedResults}

The predicted number of background events in the signal region, the number of events observed in the data, and the  predicted yield for two potential signals are presented in Table~\ref{tab:resolvedResults}.
The numbers of predicted background events and observed events are in agreement. 
\begin{table}[!ht]
\begin{center}
\begin{tabular}{ l c  }
\toprule
 Sample      & Signal Region Yield \\ 
\midrule
Multijet     & 43.3 $\pm$ 2.3   \\
\ttbar       &  ~~4.3 $\pm$ 3.0   \\
$Z$+jets     &  ~~--           \\
\midrule
Total        & 47.6 $\pm$ 3.8   \\
 \midrule
Data         & 46    \\
\midrule
SM $hh$ & ~~0.22 $\pm$ 0.05 \\
\Grav$\left(800\,\GeV\right)$, \kMPl = 1 & ~~5.7 $\pm$ 1.5 \\
\bottomrule
\end{tabular}
\caption{The number of predicted background events in the $hh$ signal region for the resolved analysis, compared to the data. The yield for two potential signals, SM nonresonant Higgs pair production and an 800\,\GeV{} \Grav resonance with \kMPl = 1 are shown.
  The quoted errors include both the statistical and systematic uncertainties.  }
\label{tab:resolvedResults} 
\end{center}
\end{table}

Figure~\ref{fig:resolvedHHUnblinded} shows a comparison of the predicted \mfourj~background distribution to that observed in the data. 
The predicted background and observed distributions are in agreement, with no significant local excesses.

\begin{figure}[!ht]
\begin{center}
\includegraphics[width=0.58\textwidth]{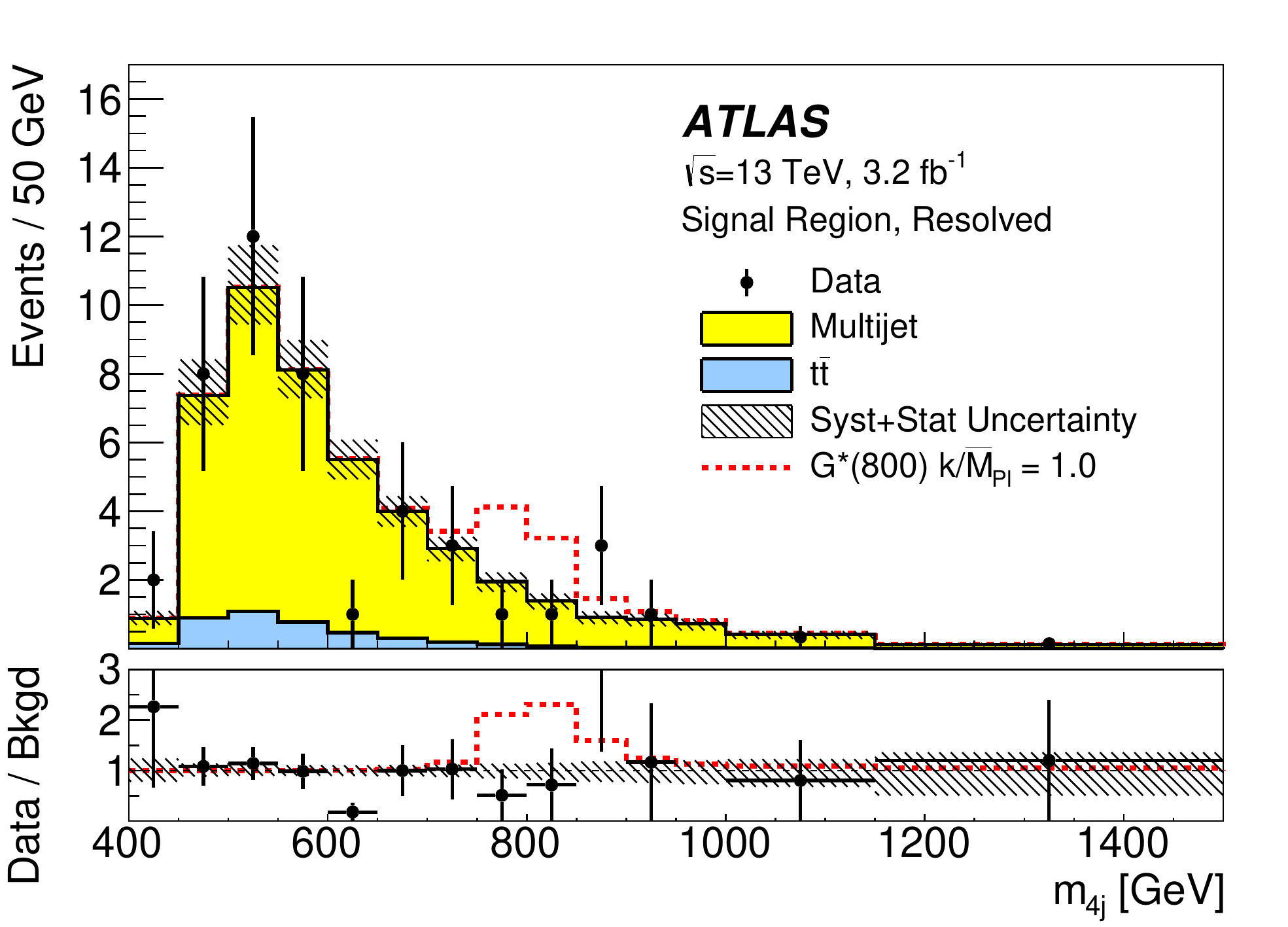}
\caption{Distribution of $m_{4j}$ in the signal region of the resolved analysis for data compared to the predicted background. 
         The hatched bands represent the combined statistical and systematic uncertainty in the total background estimate. 
         The expected signal distribution for a \Grav\ resonance with mass of 800 \GeV{} is also shown.}
\label{fig:resolvedHHUnblinded}
\end{center}
\end{figure}

\subsection{Boosted analysis}
\label{sec:boosted}

\subsubsection{Selection}
\label{sec:BoostedSel}

The boosted analysis selects events with at least two large-$R$ jets 
with 250 $<$ \pt $<$ 1500 \GeV{}, $|\eta| <$~2.0, and mass $m_{\mathrm{J}}>50$ \GeV{}.
The upper bound on the transverse momentum and the mass requirement correspond
to the kinematic region where jet calibration uncertainties are available
from Refs.~\cite{ATL-PHYS-PUB-2015-035} and \cite{Aad:2015rpa}.
Only the two large-$R$ jets with highest \pt\ are retained for further selection.
In order to reduce the contamination from \ttbar\ events,
the leading jet is additionally required to have \pt\ $>$ 350 \GeV{}, thus ensuring
that all top-quark decay products are contained in a single large-$R$ jet with
mass close to that of the top quark.

At least two track jets must be found by the ghost method~\cite{Cacciari:2007fd} to
be associated with each large-$R$ jet. 
They are required to be consistent with the primary
vertex of the event as well as to satisfy \pt\ $>$ 10 \GeV{} and $|\eta| <$ 2.5.

Since high-mass resonances tend to produce jets that are more central than
multijet background processes, the two large-$R$ jets are required to
have a separation $|\Delta \eta| <$ 1.7.

Signal event candidates are selected if each of the large-$R$ jets has a mass
consistent with that of the Higgs boson. This is defined as for the resolved analysis in Eq.~(\ref{eq:Xhh}),
where the small-$R$ dijet mass is replaced by the large-$R$ jet mass, requiring $X_{hh} <$ 1.6.
This requirement defines the signal region in the leading--subleading large-$R$ jet mass plane.

Two samples of events are selected based on the number of $b$-tagged leading
and subleading track jets associated with each large-$R$ jet.
They are referred to as the ``3-tag'' and the ``4-tag'' samples,
and require exactly three or at least four track jets passing the $b$-tagging 
selection, respectively. 
In the 3-tag sample, the fourth jet is explicitly required to fail 
the $b$-tagging requirements to define orthogonal samples.

The signal region corresponds to the kinematic requirements illustrated by the inner region 
in Figure~\ref{fig:boosted_mJ1mJ2}. The data shown in this figure
are derived from a sample of events that satisfy all selection criteria 
except for having only two track jets that pass the $b$-tagging requirements, 
referred to as the ``2-tag'' sample. This sample is used to estimate the background
contribution as described below.

\begin{figure}[!htb]
\begin{center}
\includegraphics[width=0.60\textwidth]{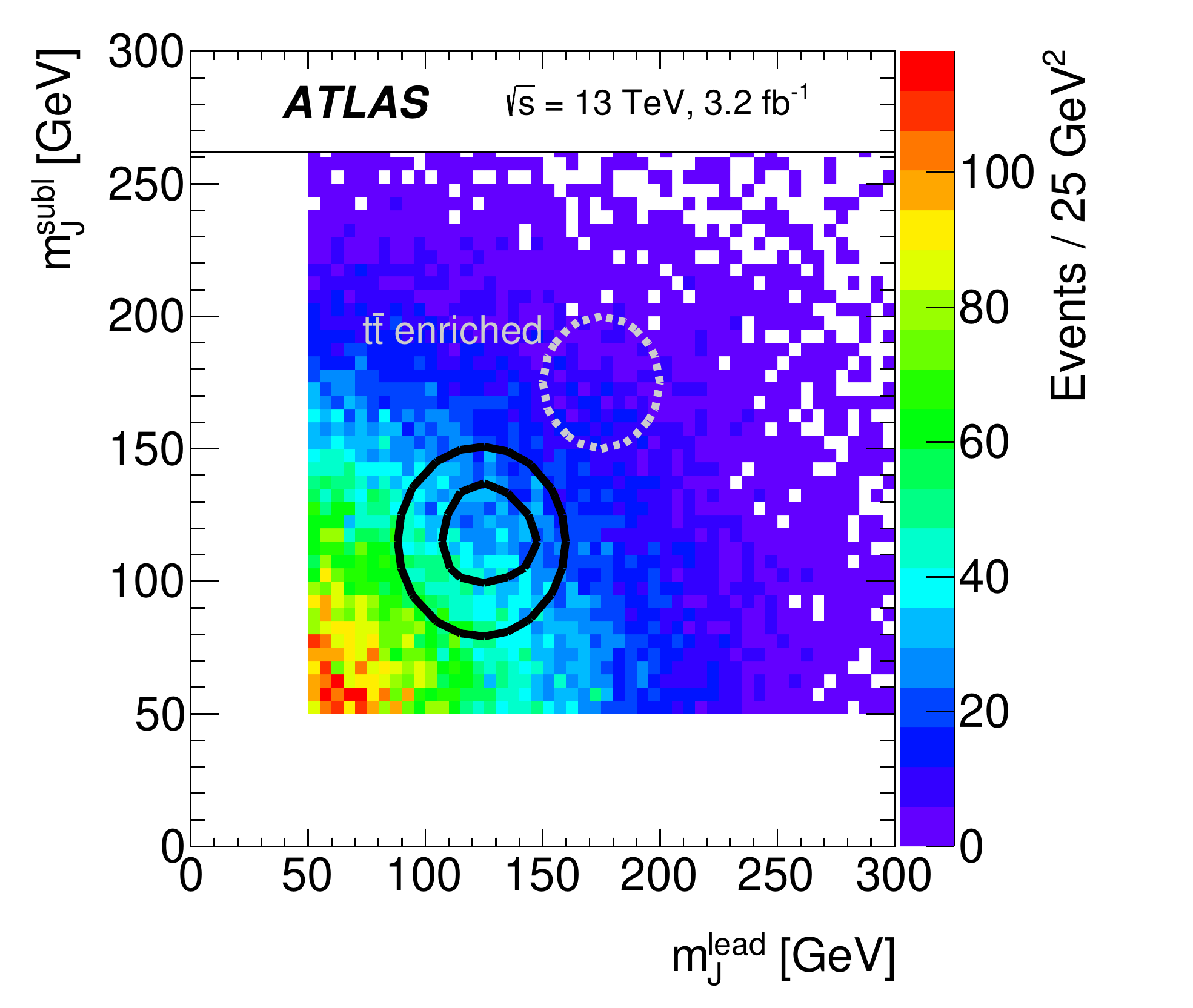}
\caption{The \msublJ\ vs. \mleadJ\ distribution for the 2-tag data sample used to model the multijet background in the boosted analysis. 
  The signal region is the area surrounded by the inner black contour line,
  centered on $\mleadJ =124$\ \GeV, $\msublJ =115$\ \GeV. 
        The control region is the area inside the outer black contour line, excluding the signal region. 
        The sideband region is the area outside the outer contour line.
        The kinematic region enriched in $t\bar{t}$ events is indicated by the
  dashed white contour line.}
\label{fig:boosted_mJ1mJ2}
\end{center}
\end{figure}

\begin{figure}[!htb]
\begin{center}
\includegraphics[width=0.45\textwidth]{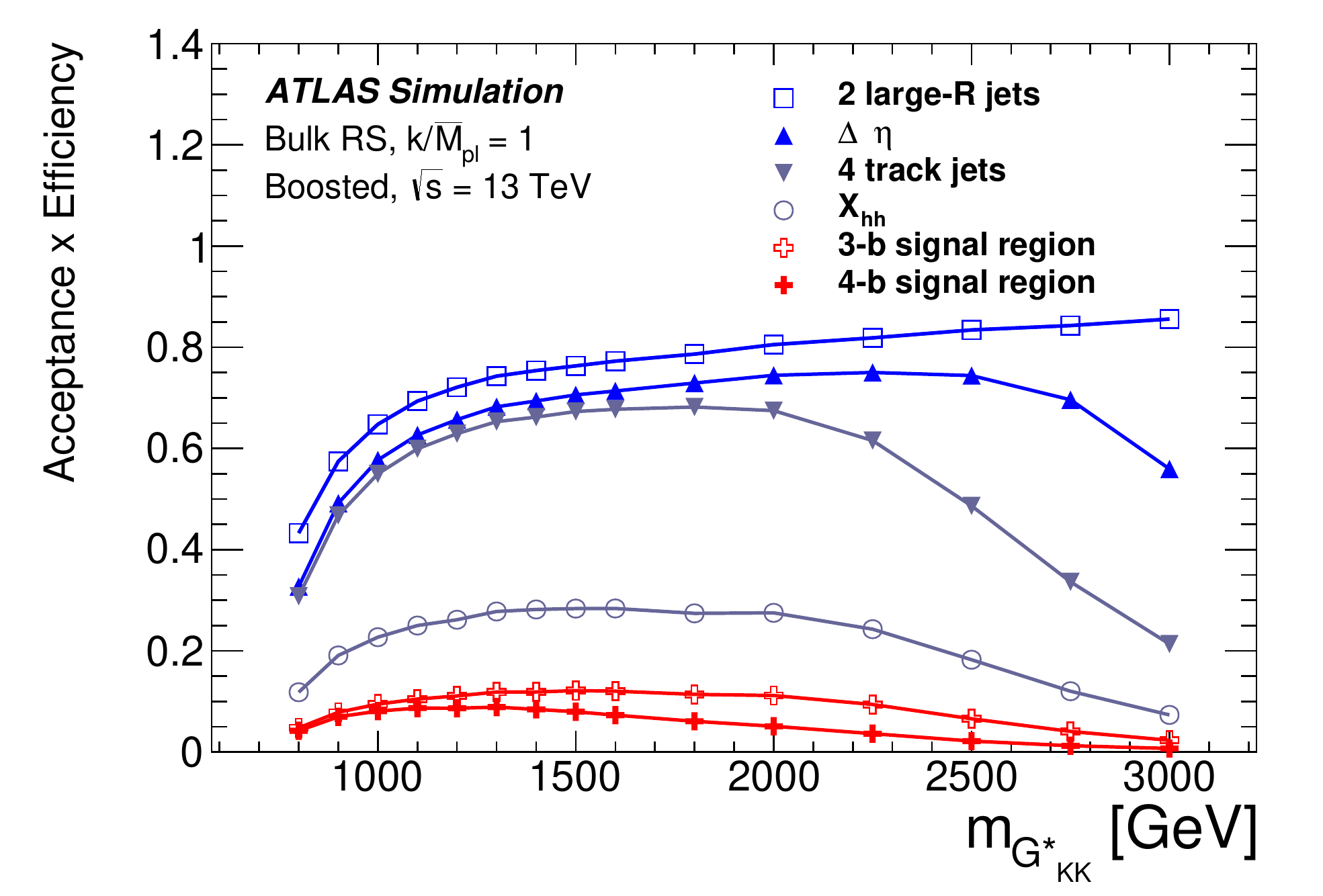}
\includegraphics[width=0.45\textwidth]{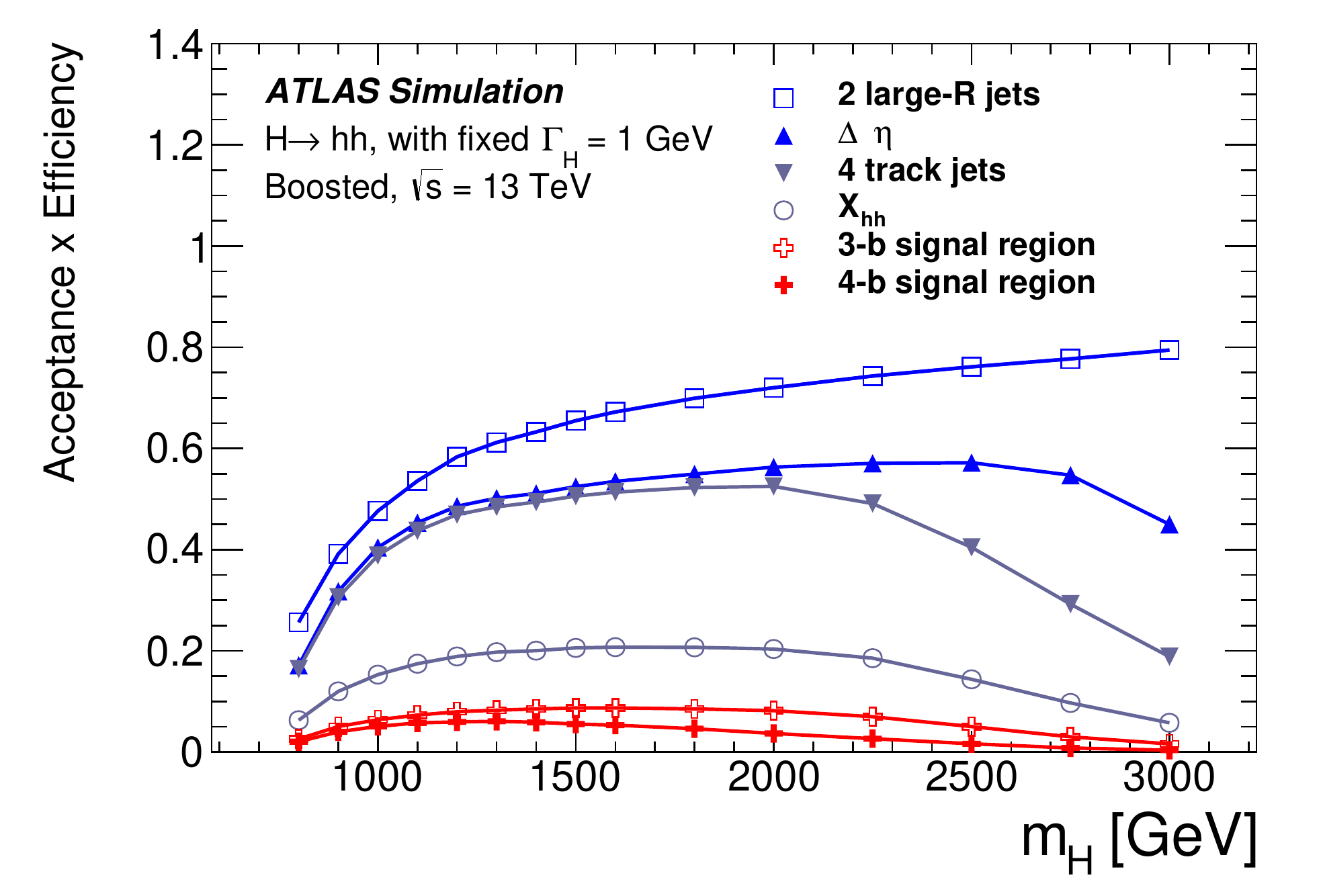}
\caption{The selection efficiency as a function of resonance mass at each stage of the event selection for (left) \Gtohhb and (right) \Htohhb decays in the boosted analysis. 
The ``2 large-$R$ Jets'' curve includes the requirements that the leading jet has $\pt\ > 350 \GeV$ and that the large-$R$ jets have a mass greater than 50 \GeV. }
\label{fig:cutflowPlotsBoosted}
\end{center}
\end{figure}

The acceptance times efficiency for each stage of the boosted-analysis event selection is shown in Figure \ref{fig:cutflowPlotsBoosted} for the \Grav and heavy scalar models.
The requirement that at least two individual track jets be associated to the large-$R$ jets becomes less efficient at high mass due to merging. 
The full selection for a \Grav\ resonance with a mass of 1000 \GeV{}~(2000 \GeV{}) and $\kMPl = 1$ has an acceptance times efficiency of $9$\%~($11$\%) in the 3-tag sample and $8$\%~($5$\%) in the 4-tag sample.

\subsubsection{Background estimation}
\label{sec:BoostedBkg}

As in the resolved analysis, the dominant source of background stems from
multijet (80--90\%) events and the rest is primarily due to \ttbar\ production. 
The background estimation method generally follows the same approach as that
described in Section~\ref{sec:ResolvedBkg}. Differences are highlighted below.

The shape of the multijet background in both the 3-tag and 4-tag samples
is derived from the 2-tag sample.
Due to the large statistical uncertainty in the background prediction for dijet
masses (\mtwoJ) above 1500 \GeV,
an exponential fit to the data in the range between 900 and 2000 \GeV{}
is used to model the high-mass tail of the dijet distribution in the signal region. 
The estimated background yield in each signal region, $N^{\text{3(4)-tag}}_{\text{bkg}}$, is 
computed according to

\begin{equation}
  N^{\text{3(4)-tag}}_{\text{bkg}} = \muqcd^{\text{3(4)-tag}}\, N^{\text{2-tag}}_{\text{Multijet}}
  + \alpha_{t\bar{t}}^{\text{3(4)-tag}}\, N^{\text{3(4)-tag}}_{t\bar{t}}
  + N^{\text{3(4)-tag}}_{Z+\text{jets}}~,
\label{eqn:BoostedBackground}
\end{equation}

\noindent
where $N^{\text{2-tag}}_{\text{Multijet}}$ is the number of multijet events in the 2-tag sample,
$N^{\text{3(4)-tag}}_{t\bar{t}}$ and $N^{\text{3(4)-tag}}_Z$ are the numbers of events predicted by
the 3(4)-tag \ttbar\ and $Z$+jets MC samples.
The parameter $\muqcd^{\text{3(4)-tag}}$ corresponds to the ratio
of multijet event yields in the 3(4)-tag and 2-tag samples, as defined in 
Eq.~(\ref{eqn:qcdnorm}), except for considering 3- or 4-tag events in the numerator.
Finally, the parameter $\alpha_{t\bar{t}}^{\text{3(4)-tag}}$ is a scale factor designed to correct the \ttbar\
event yield estimated from the MC simulation.

A sideband region defined by $\sqrt{(\mleadJ - 124 \GeV{})^2  + (\msublJ - 115 \GeV{})^2} > 36$ \GeV{}
is used to measure $\muqcd^{\text{3(4)-tag}}$ and $\alpha_{t\bar{t}}^{\text{3(4)-tag}}$ from the data.
The background estimate is validated in a control region defined to be complementary to the
sideband and signal regions.

\begin{figure}[!ht]
\begin{center}
\includegraphics[width=0.49\textwidth]{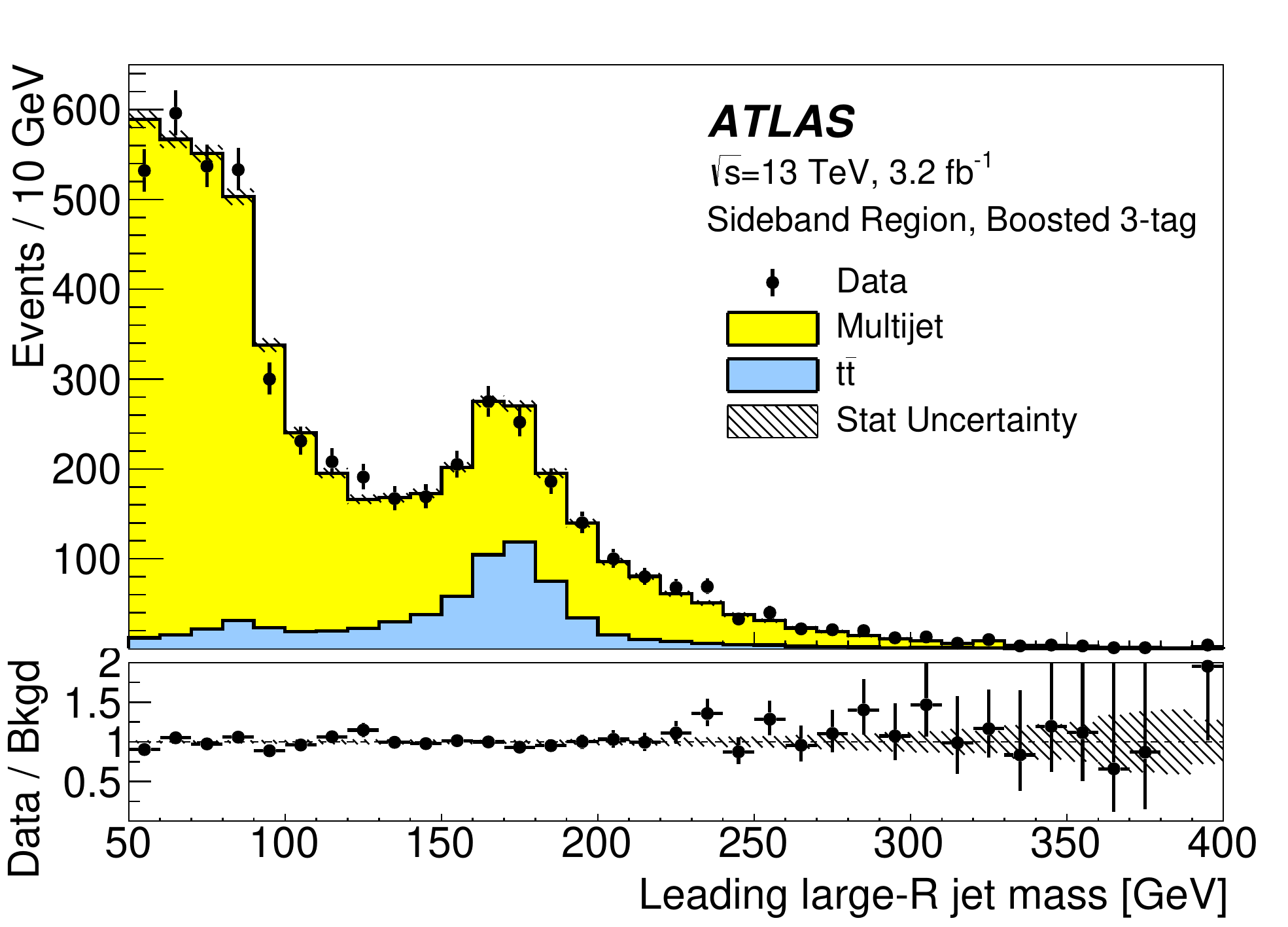}
\includegraphics[width=0.49\textwidth]{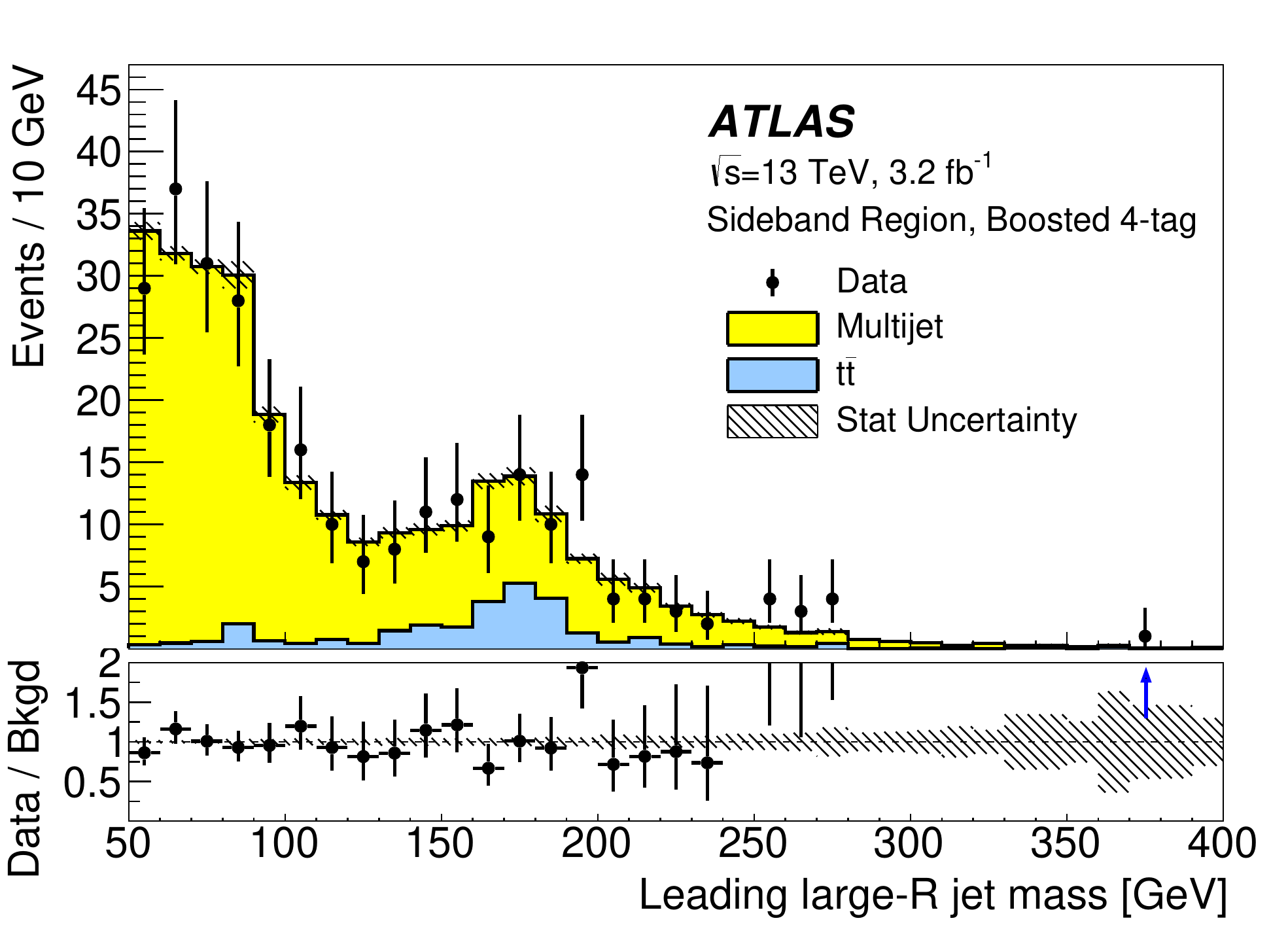}
\end{center}
\caption{
  The leading large-$R$ jet mass distribution in the $hh$ sideband region for data (points) and background
  estimate (histograms) in the boosted analysis for events in the (left) 3-tag and (right) 4-tag categories.
  The shape of the multijet distributions is taken from the 2-tag region and is fitted to the data.
  The hatched bands represent the statistical uncertainty in the total background estimate.
}
\label{fig:LeadJMass_SB_boosted}
\end{figure}

Both $\muqcd^{\text{3(4)-tag}}$ and $\alpha_{t\bar{t}}^{\text{3(4)-tag}}$ are extracted from a binned likelihood fit to the
leading large-$R$ jet mass distribution obtained in the sideband region of the 3(4)-tag sample, as shown in Figure~\ref{fig:LeadJMass_SB_boosted}.
In this fit, the multijet distribution is extracted from the
2-tag sample, after subtraction of the \ttbar\ and $Z$+jets contributions
predicted by the MC simulation.
The \ttbar\ and $Z$+jets distributions in the sideband region of the 3(4)-tag sample
are taken  from the MC simulation.
The resulting fit values and their statistical uncertainties
for the 3-tag sample are $\muqcd^{\text{3-tag}} = 0.160 \pm 0.003$
and $\alpha_{t\bar{t}}^{\text{3-tag}} = 1.02 \pm 0.09$, with a correlation 
coefficient of $-0.60$ between these two parameters. 
The corresponding values measured in the 4-tag sample are
$\muqcd^{\text{4-tag}} = 0.0091 \pm 0.0007$ and $\alpha_{t\bar{t}}^{\text{4-tag}} = 0.82 \pm 0.39$,
with a correlation coefficient of $-0.58$. 
A large anti-correlation is observed since the multijet and \ttbar\ background
contributions are constrained to add up to the total number of events in the sideband
region of the 3-tag and 4-tag data samples.

The modeling of the background yield and kinematics is validated in the control
region of the 3-tag and 4-tag samples.
Good agreement is observed between the data and the predicted background in both the sideband
and control regions of the 3-tag and 4-tag samples as shown in Table~\ref{tab:BoostedSBCR}.
The shapes of the \ttbar\ kinematic distributions in the 4-tag signal region
are extracted from the MC simulation in the 3-tag signal region due to the limited size of the 4-tag MC sample.

\begin{table}[!ht]
\begin{center}
\begin{tabular}{ l c  c  }
\toprule
 Sample (3-tag) & Sideband Region       & Control Region  \\ 
\midrule
Multijet        & 4328   $\pm$ 27       & 607  $\pm$  10  \\
\ttbar          & 683.5  $\pm$ 8.1      & 99.6 $\pm$  3.1 \\
$Z$+jets        & ~~31.8   $\pm$ 3.7      & ~~7.7  $\pm$  1.8 \\
\midrule
Total           & 5043 $\pm$ 28     & 715 $\pm$  11 \\
\midrule
Data            & 5043                  & 724     \\	
\toprule
  Sample (4-tag) & Sideband Region      & Control Region  \\ 
\midrule
Multijet         & 247.4 $\pm$ 1.5      & 34.7  $\pm$  0.6 \\
\ttbar           & ~~28.4  $\pm$ 1.5      & ~~5.1   $\pm$  0.7 \\
$Z$+jets         & ~~~~3.4   $\pm$ 1.2      & ~~0.6   $\pm$  0.5  \\
\midrule
Total            & 279.2 $\pm$ 2.5      & 40.3  $\pm$  1.0  \\
\midrule
Data             & 279                  & 45     \\	
\bottomrule
\end{tabular}
\caption{The number of events in data and predicted background events in the $hh$ sideband and control
  regions of the 3-tag and 4-tag samples for the boosted analysis.
  The number of multijet and $t\bar{t}$ background events in the sideband regions
  are constrained by the number of observed events, as explained in the text.  
  The uncertainties are purely statistical.}
\label{tab:BoostedSBCR} 
\end{center}
\end{table}

\subsubsection{Systematic uncertainties}
\label{sec:BoostedSyst}

Evaluation of systematic uncertainties in the boosted analysis generally follows the same 
approach as that described in Section~\ref{sec:ResolvedSyst}. Differences are highlighted
here.

The large-$R$ jet energy resolution and scale uncertainties 
as well as the jet mass resolution (JMR)
and scale (JMS) uncertainties are derived in situ
from 8 \TeV{} $pp$ collisions, taking into account MC simulation extrapolations for the
different detector and beam conditions present in 8 and 13 \TeV{} data-taking periods
\cite{ATL-PHYS-PUB-2015-033}.
The uncertainty in the $b$-tagging efficiency for track jets is evaluated with the
same method used for $R = 0.4$ calorimeter jets.

Systematic uncertainties in the normalization and shape of the background model are assessed in the control region. 
The background predictions in both the 3-tag and 4-tag control regions agree with the observed data to within statistical uncertainties.
The statistical uncertainties in the control region yields are assigned
as systematic uncertainties in the multijet background normalization. 
The uncertainty in the shape of the multijet background is assessed in the
control region via a linear fit to the ratio of the distributions shown 
in Figure~\ref{fig:m2J_CR_boosted}.

\begin{figure}[htbp!]
\begin{center}
\includegraphics[scale=0.4]{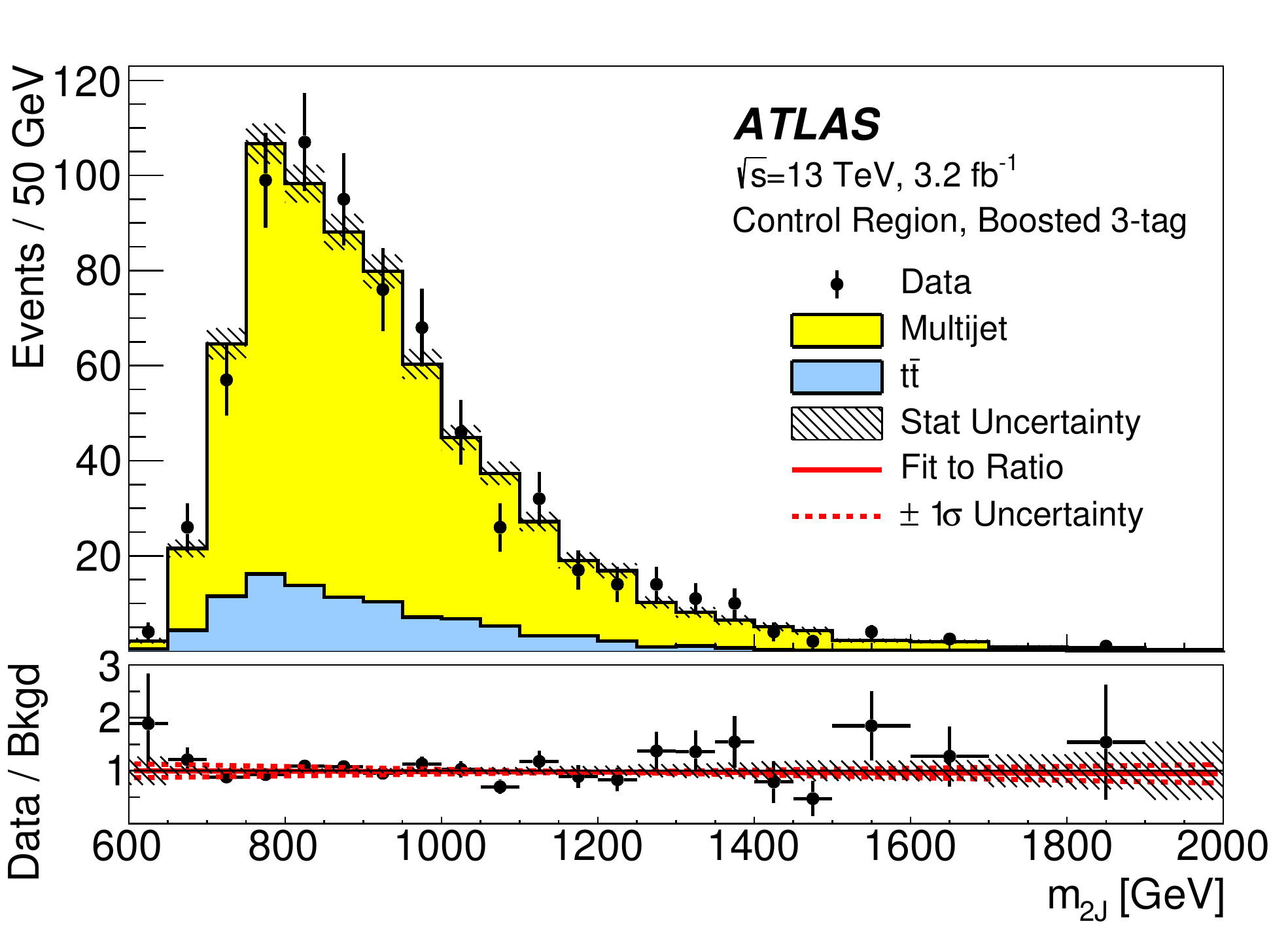}
\includegraphics[scale=0.4]{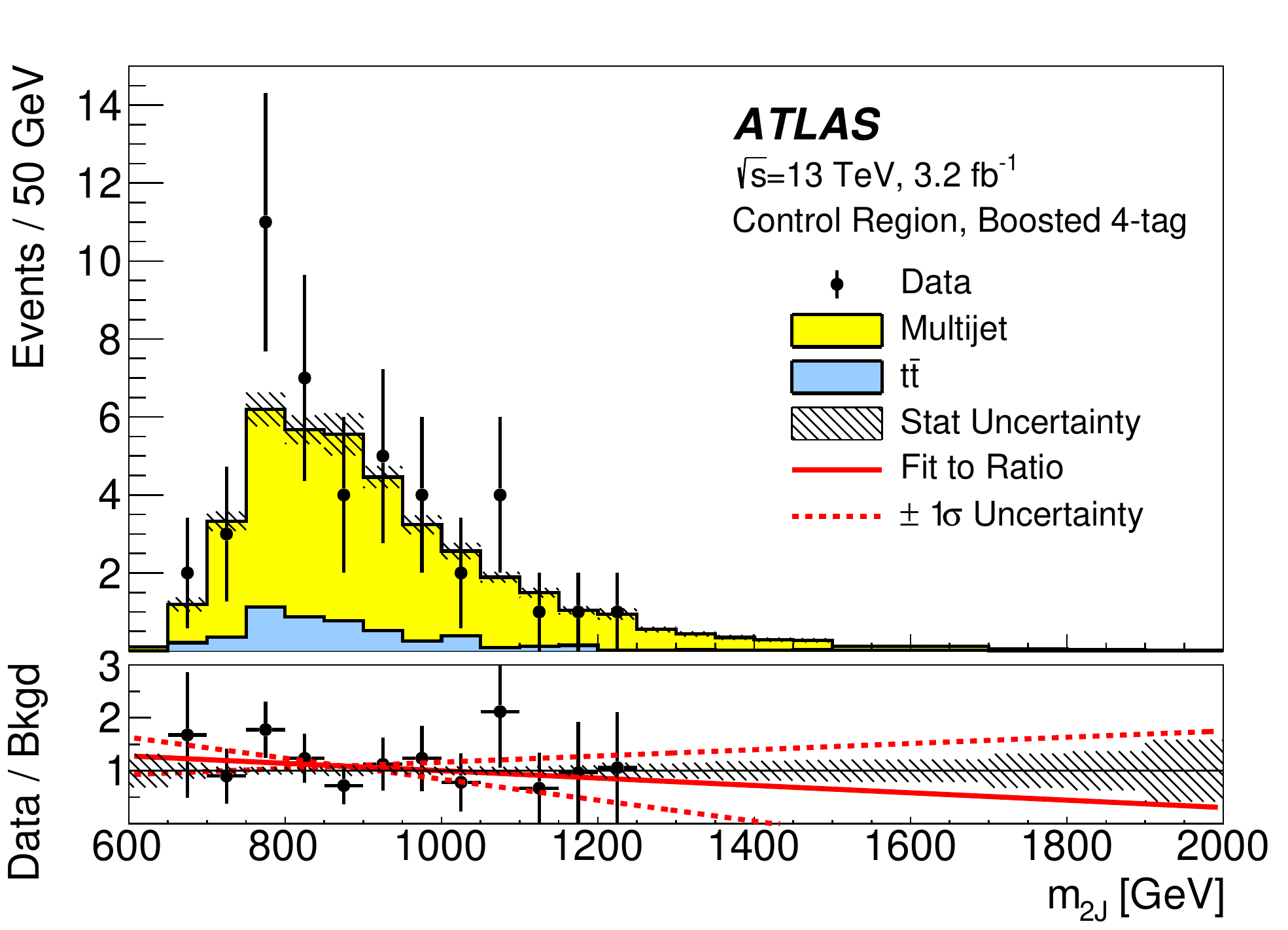}
\end{center}
\caption{Dijet mass distribution in the control region for data (points) and 
 background estimate (histograms) in the boosted analysis for events in the
 (left) 3-tag and (right) 4-tag categories. 
 The hatched bands represent the statistical uncertainty in the total background estimate. 
}
\label{fig:m2J_CR_boosted}
\end{figure}

An additional uncertainty in the shape of the tail of the background prediction
is assigned by fitting the 2-tag dijet mass distribution with a variety of empirical functions
designed to model power-law behavior, as described in Ref.~\cite{Aad:2015mzg}. 
The largest difference between the exponential function predictions and those
from alternative fit functions,
considering the variation of the fitted parameters within their statistical uncertainties,
is taken as a systematic uncertainty.

Relative systematic uncertainties in both the background and signal event yields are summarized
in Table~\ref{tab:BoostedSyst} for the 3-tag and 4-tag selections.
For the background, the entry labeled ``Statistical'' corresponds to the
statistical uncertainty from the fit to the leading large-$R$ jet mass (see
Section~\ref{sec:BoostedBkg}) used to extract the multijet and \ttbar\ background
yields, taking the correlation between these yields into account. It also includes
the \ttbar\ modeling uncertainties and the statistical uncertainty associated 
with the data yield in the 2-tag sample.
Uncertainties in the \mtwoJ shape of the multijet and \ttbar backgrounds are 
not listed in Table~\ref{tab:BoostedSyst},
as they do not affect the event yields, 
but are accounted for in the statistical analysis.

\begin{table}[ht!]
\newcommand\T{\rule{0pt}{2.6ex}}
\begin{center} 
\begin{tabular}{  l  c  c  c  c }
\toprule
 Source          & Background    & \multicolumn{2}{c}{\Grav (1500 \GeV)} & $H$ (1500 \GeV) \\
                 &         & \kMPl = 1 & \kMPl = 2  &     \\
\midrule
   Luminosity    &  --     &  5.0    &  5.0    & 5.0 \\
\midrule
                \multicolumn{5}{c}{3-tag}  \\
\midrule
   JER           &  $<1$   &  $< 1$   &  $<1$   & $< 1$ \\
   JES           &  ~~$2$    &  $<1$    &  $< 1$  & $<1$ \\
   JMR           &  ~~$1$    &  $12$    &  $12$   & $11$  \\
   JMS           &  ~~$5$    &  $14$    &  $13$   & $17$  \\
   $b$-tagging   &  ~~$1$    &  $23$    &  $22$   & $23$  \\
   Theoretical   &  ~~--      &  ~~$3$     &  ~~$3$    & ~~$3$ \\
   Multijet      &  ~~$3$    &  ~~--       &  ~~--      & ~~--  \\
   Statistical   &  ~~$2$    &  ~~$1$     &  ~~$1$    &  ~~$1$  \\
   \midrule
   Total         &  ~~$7$    &  $31$   &  $30$   &  $33$ \\
\bottomrule
                \multicolumn{5}{c}{\T 4-tag}  \\
\midrule
   JER           &  $< 1$    &  $<1$  &  $<1$   & $<1$ \\
   JES           &  $< 1$    &  $<1$    &  $<1$    & $<1$ \\
   JMR           &  ~~$4$    &  $12$   &  $13$   & $13$  \\
   JMS           &  ~~$5$    &  $13$   &  $13$   & $14$  \\
   $b$-tagging   &  ~~$2$    &  $36$   &  $36$   & $36$  \\
   Theoretical   &  ~~--     &  ~~$3$    &  ~~$3$    & ~~$3$ \\
   Multijet      &  $14$  &  ~~--      &  ~~--      &  ~~--  \\
   Statistical    &  ~~$3$    &  ~~$1$      &  ~~$1$      &  ~~$1$  \\
   \midrule
   Total         &  $15$   &  $42$   &  $42$   & $43$ \\
\bottomrule
\end{tabular}
\caption{Summary of systematic uncertainties (expressed in percentage yield) 
 in the total background and signal event yields in the 3-tag and 4-tag signal regions in the boosted analysis.
 Uncertainties are provided for a \Grav resonance mass of 1500 \GeV{}
 with \kMPl = 1 or 2, as well as for a spin-0 narrow-width $H$ boson.
}
\label{tab:BoostedSyst}
\end{center}
\end{table}

\subsubsection{Event yields}

The predicted number of background events in the 3-tag and 4-tag signal regions,
the number of events observed in the data, and the predicted yield for a potential signal
are reported in Table~\ref{tab:BoostedResults}.
One event in the 4-tag signal region, with a mass of 852 \GeV, is in common with the resolved analysis.
The dijet mass distribution in the signal region is shown in Figure~\ref{fig:m2J_SR_boosted}.
An excess of data is observed in the 3-tag signal region for $\mtwoJ \sim 900 \GeV{}$ and in the range
between 1600 and 2000 \GeV{}. The significance of these excesses is evaluated below.

\begin{table}[!htb]
\begin{center}
\begin{tabular}{ l c c }
\toprule
 Sample  & Signal Region (3-tag)    & Signal Region (4-tag) \\ 
\midrule
Multijet &  235  $\pm$ 14       & 13.5 $\pm$ 2.4 \\
\ttbar 	 &  ~~48   $\pm$ 22       & ~~1.2  $\pm$ 1.0 \\
$Z$+jets &  ~~2.0  $\pm$ 2.2      &       ~~--        \\
\midrule
Total 	 &  285  $\pm$ 19       & 14.6 $\pm$ 2.4 \\
 \midrule
Data 	 &  316                 & 20 \\
\midrule
\Grav$\left(1000~\GeV{}\right)$, \kMPl = 1 & ~~3.4 $\pm$ 0.9  & ~~2.9 $\pm$ 1.1 \\
\bottomrule
\end{tabular}
\caption{The number of predicted background
  events in the $hh$ 3-tag and 4-tag signal regions, compared to the data for the boosted analysis.
  Errors correspond to the total uncertainties in the predicted event yields.
  The yields for a 1000~\GeV{} \Grav in 
  the bulk RS model with \kMPl = 1 is also given. }
\label{tab:BoostedResults} 
\end{center}
\end{table}

\begin{figure}[htbp!]
\begin{center}
\includegraphics[scale=0.4]{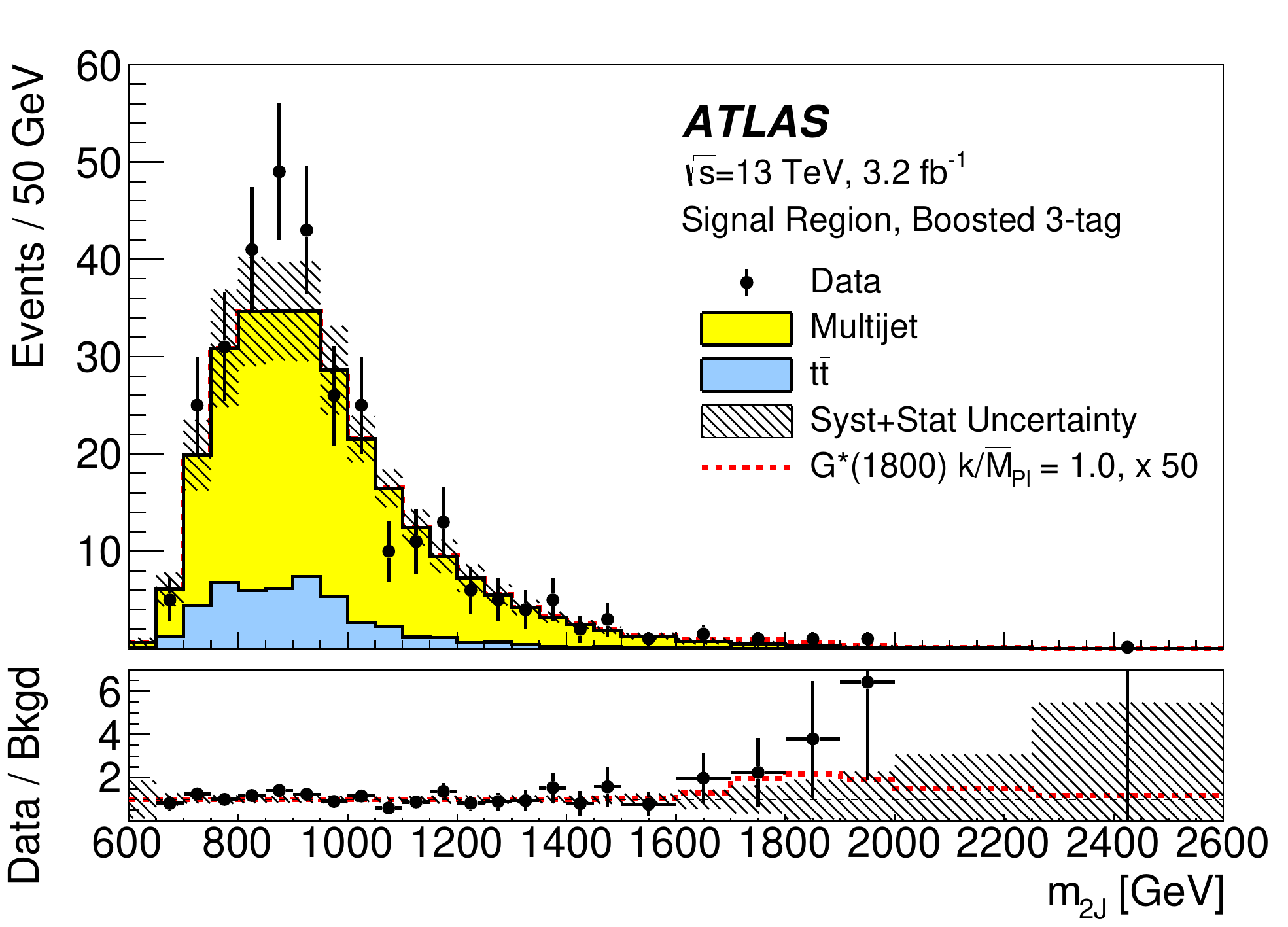}
\includegraphics[scale=0.4]{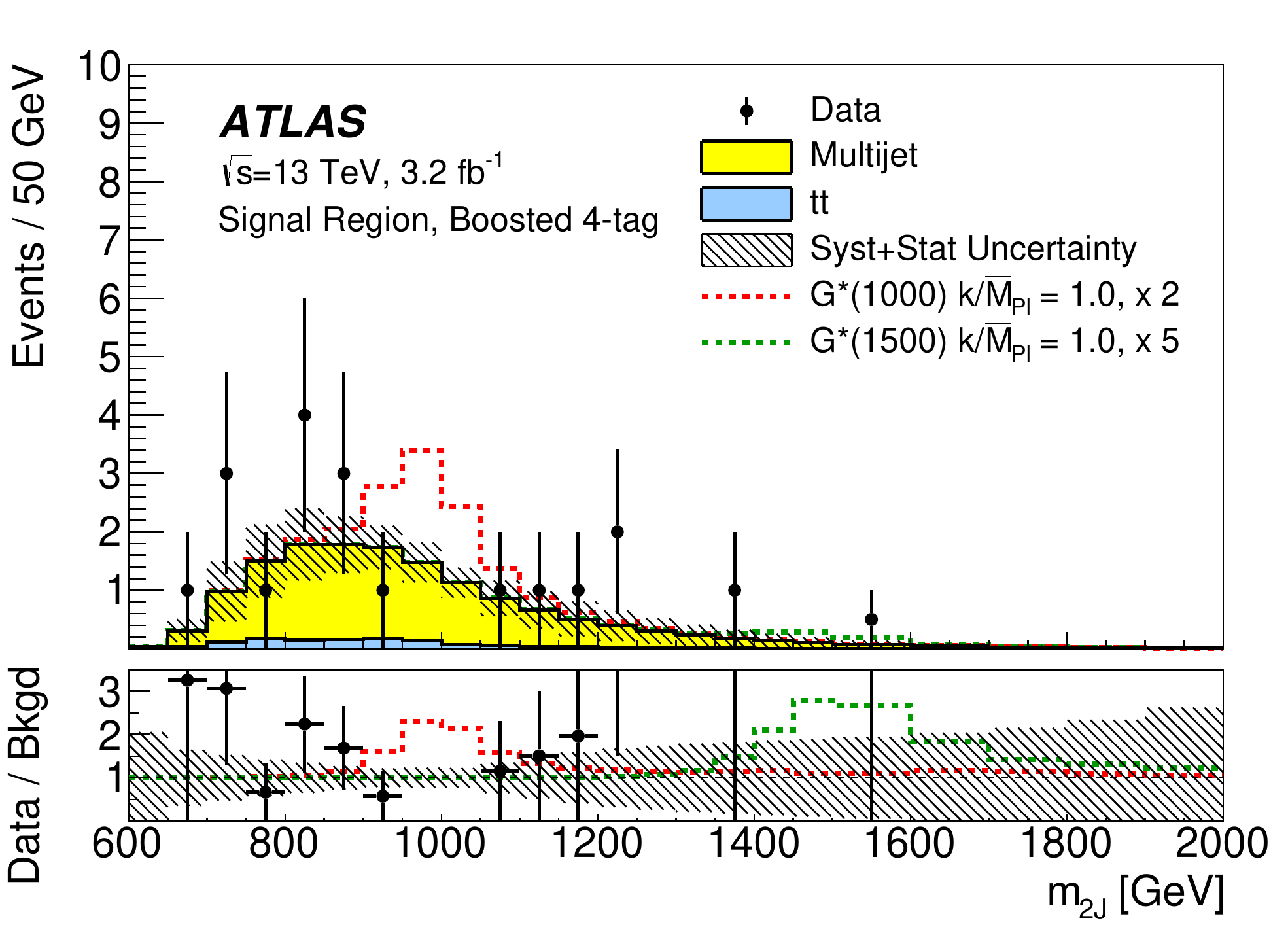}
\end{center}
\caption{Dijet mass distribution in the $hh$ signal region for data (points) and 
 background estimate (histograms) in the boosted analysis for events in the
 (left) 3-tag and (right) 4-tag categories. The expected signal
 distributions for \Grav\ masses of 1000, 1500 and 1800 \GeV{} are also shown.
 The uncertainty band includes both the statistical and systematic uncertainties in the background estimate.}
\label{fig:m2J_SR_boosted}
\end{figure}

\section{Results}
\label{sec:results}
The results from the resolved and boosted analyses are interpreted separately
using the statistical procedure described in Ref.~\cite{Aad:2012tfa} and references therein.
A test statistic based on the profile likelihood ratio~\cite{Cowan:2010js} is used
to test hypothesized values of $\mu$, the global signal strength factor, separately for each model tested.
The statistical analysis described below is performed using the data observed in the signal regions.
The systematic uncertainties are treated as independent within each signal region using
Gaussian or log-normal constraint terms in the definition of the likelihood function.
In the boosted analysis, the data from the 3-tag and 4-tag signal regions are fitted simultaneously 
treating data-derived systematic uncertainties related to the multijet background
estimate as uncorrelated and all other systematic uncertainties as fully correlated.
In the case of the search for nonresonant $hh$ production, only the number of events passing
the final selection is used whereas the $m_{\mathrm{4j}}$ or $m_{\mathrm{2J}}$ distributions are used
in the case of the search for $hh$ resonances.

\subsection{Background-only hypothesis tests}

In order to determine if there are any statistically significant local excesses in the data,
a test of the background-only hypothesis ($\mu=0$) is performed.
The significance of an excess is quantified using the local $p_{0}$, the probability that the
background could produce a fluctuation greater than or equal to the excess observed in data.
A global $p_0$ is also calculated for the most significant discrepancy, using background-only
pseudoexperiments to derive a correction for the look-elsewhere effect across the mass range tested~\cite{Gross:2010qma}.

In the case of the resolved analysis,
the largest deviation from the background-only hypothesis occurs around 900 \GeV{} 
and is found to have a local significance less than $2\,\sigma$.

In the case of the boosted analysis, the largest local deviation corresponds to a broad data excess 
in the 3-tag signal region starting at $\mtwoJ\sim 1700$~\GeV{}.
The local significance of this excess is $2.0\,\sigma$ assuming a \Grav resonance with \kMPl = 1. 

\subsection{Exclusion limits}

The data are used to set upper limits on the cross sections for the different benchmark signal processes.
Exclusion limits are based on the value of the statistic \cls~\cite{Read:2002hq}, with a value of $\mu$
regarded as excluded at the 95\% confidence level (CL) when \cls~is less than 5\%. 

The nonresonant search is performed using the resolved analysis, since it has better sensitivity than
the boosted analysis. Using the SM $hh$ nonresonant production as the signal model, the observed 95\% CL upper limit
is $\sigma(pp\rightarrow hh\rightarrow b\bar{b}b\bar{b}) < 1.22$ pb, 
a value to be compared with the inclusive SM prediction (as defined in Section~\ref{sec:data-mc})
of $\sigma(pp\hspace{-0.4mm}\rightarrow\hspace{-0.4mm}hh\hspace{-0.4mm}\rightarrow\hspace{-0.4mm}b\bar{b}b\bar{b}) = 11.3^{+0.9}_{-1.0}$ fb.

For the resonant Higgs-boson pair production search, the resolved and boosted analyses offer their best sensitivity in
complementary resonance mass regions.
The resolved analysis gives a more stringent expected exclusion limit for resonance masses up to
(and including) 1100 \GeV{}, while the boosted analysis offers better sensitivity beyond that mass.
A simple combination of the separate exclusion limits from the resolved and boosted analyses is used.
This is achieved by taking the limit from the analysis with the more stringent expected exclusion at each mass point for each of the signal models.

Figure~\ref{fig:CombinedLimits} shows the combined 95\% CL upper limits for three 
different resonances: a spin-2 \Grav\ in the bulk RS model with \kMPl = 1 and 2,
and a spin-0 narrow-width $H$ boson.
For the spin-2 \Grav\ with \kMPl = 1, limits on \sigGfourb\ are set in the
range between 21 and 73 fb for masses between 600 and 3000 \GeV{}.
The corresponding range of limits for the \Grav\ resonance with \kMPl = 2 is
34 to 86 fb.
Although no events are observed at masses near 3000 \GeV{}, the observed limit
remains about $1\,\sigma$ weaker than the expected limit due to a substantial
low-mass tail in the shape of high-mass resonance signals and the slight data excess 
observed at high mass.
The cross-section limits for resonance masses below 600 \GeV{} weaken substantially
due to the lower acceptance times efficiency (see Figure~\ref{fig:cutflowPlotsResolved}) 
and the increased level of background.
These cross-section upper limits translate into observed (expected) excluded mass 
ranges of 480--770 (470--735) \GeV{} for $\kMPl = 1$ and $< 965$ ($< 995$) \GeV{}
for $\kMPl = 2$.
The cross-section upper limits for the spin-0 narrow-width $H$ boson are similar, with 95\% CL exclusion limits ranging from 30 to 300 fb in the mass range between 500 and 3000~\GeV{}.

\begin{figure}[!ht]
\begin{center}
\subfloat[Bulk RS, \kMPl = 1]{\includegraphics[width=0.6\textwidth]{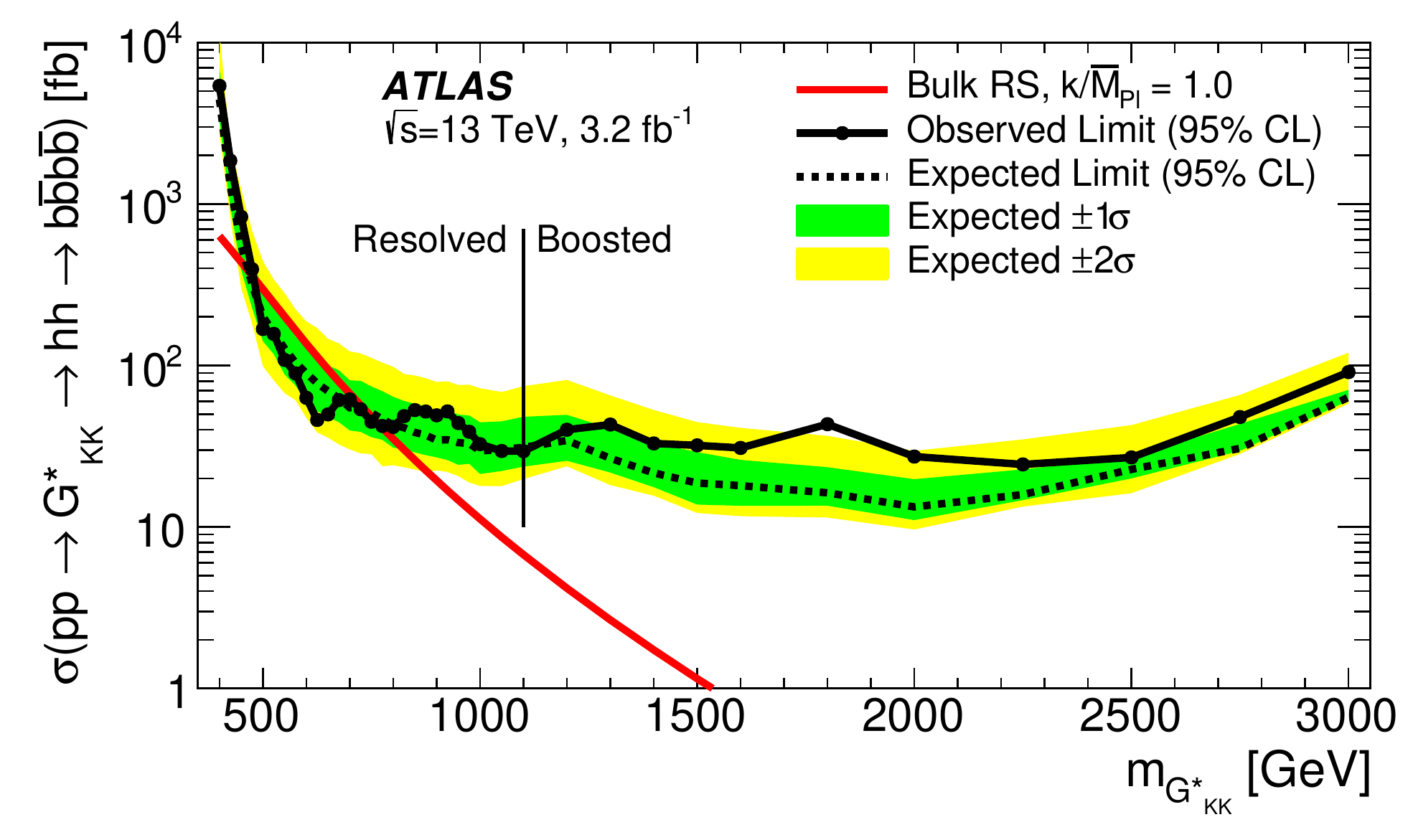}}\\
\subfloat[Bulk RS, \kMPl = 2]{\includegraphics[width=0.6\textwidth]{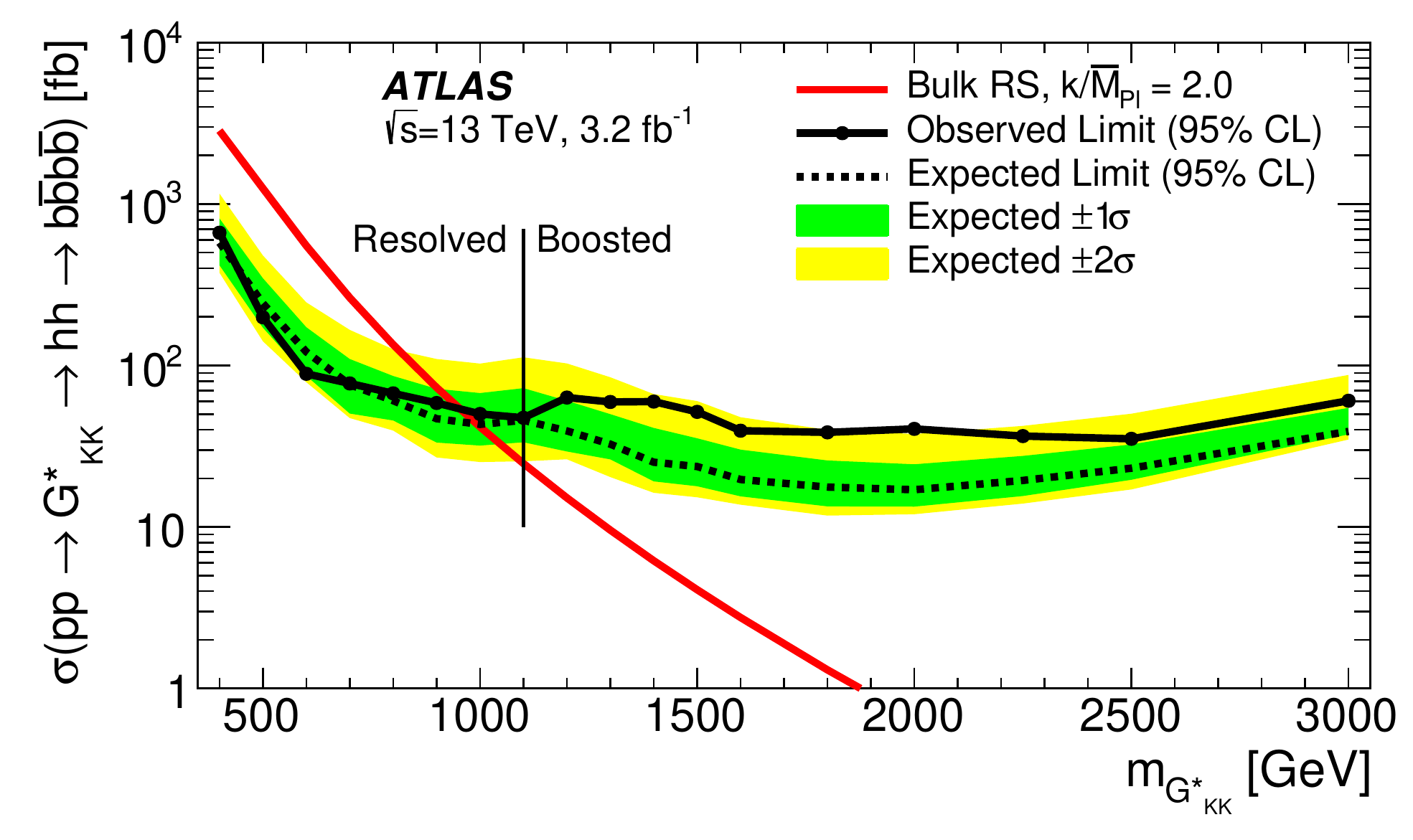}}\\
\subfloat[Spin-0 narrow-width $H$ boson]{\includegraphics[width=0.6\textwidth]{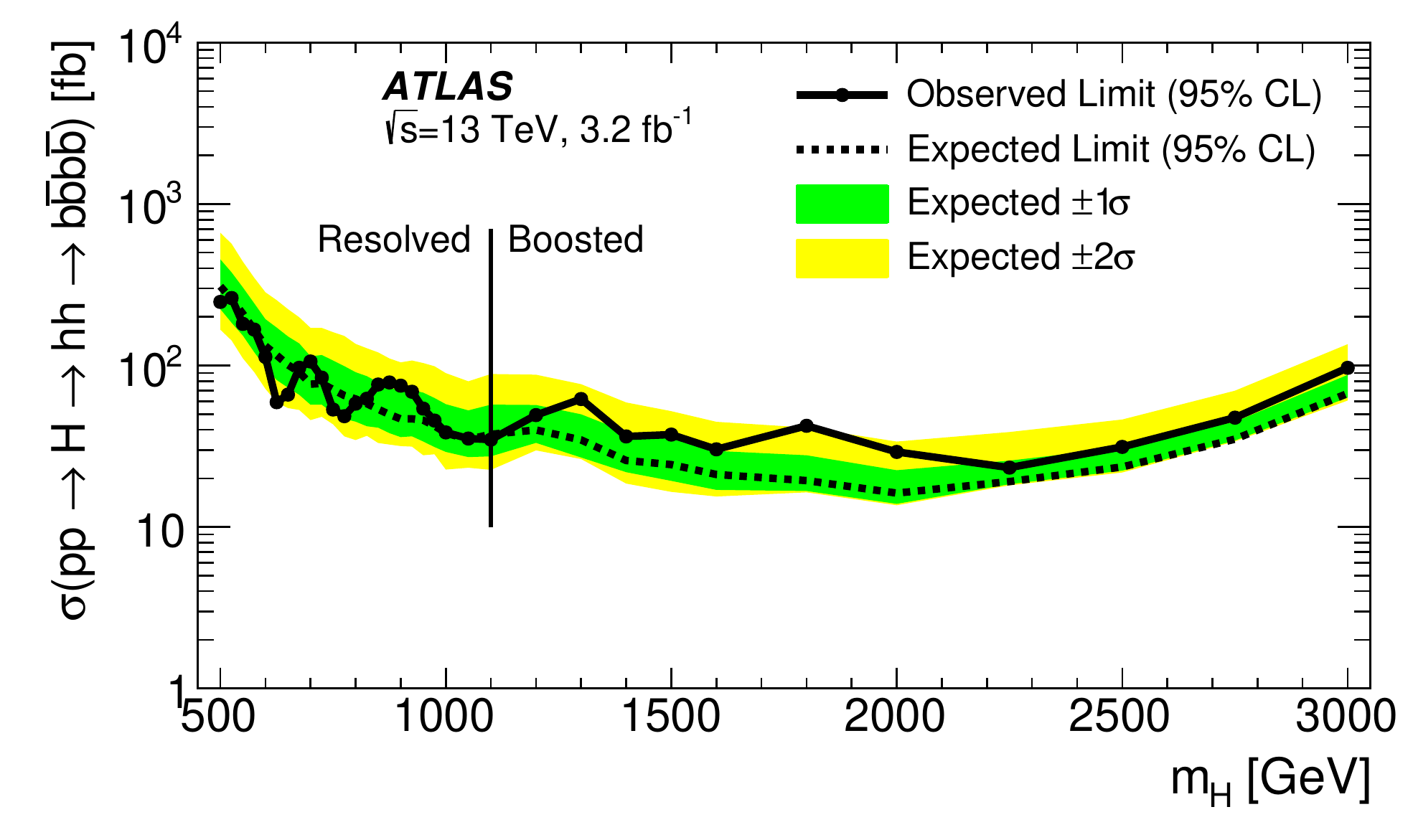}}
\caption{The expected and observed upper limit for \pptoGtofourb in the bulk RS model with (a) \kMPl = 1
  and (b) \kMPl = 2, as well as (c) \pptoHtofourb with fixed $\Gamma_H = 1\,\GeV{}$, at the 95\% confidence level.
The results of the resolved analysis are used up to a mass of 1100 \GeV{} and those of the boosted analysis are used at higher mass where its expected sensitivity is higher. The red curves show the predicted cross sections as a function of resonance mass for the models considered.
}
\label{fig:CombinedLimits}
\end{center}
\end{figure}

The search sensitivity of this analysis is similar to that achieved
at $\rts = 8$ \TeV{} with 19.5~\ifb\ for resonance masses below 1350 \GeV{} 
but exceeds it above that mass by factors of 1.4 at 1500 \GeV{} and 12 at 2000 \GeV{}.
The search has also been extended to resonance masses beyond 2000 \GeV{}, 
up to 3000 \GeV{}.

\FloatBarrier

\section{Conclusions}
\label{sec:conclusion}

A search for both resonant and nonresonant production of pairs of Standard Model Higgs bosons has been
carried out in the dominant \fourb\ channel with 3.2 \ifb\ of
$pp$ collision data collected by ATLAS during the 2015 run of the LHC
at $\rts = 13$ \TeV{}.
Results are reported for the resolved analysis with each $h \to b\bar{b}$ decay
reconstructed as two separate $b$-tagged jets and for the boosted
analysis with each $h \to b\bar{b}$ decay reconstructed as a single large-radius
jet associated with two small-radius track jets and a minimum of three $b$-tags for
the $hh$ system.
No significant data excess is observed above the estimated background
consisting mainly of multijet and \ttbar\ events.
Upper limits on the production cross section times branching ratio to the
\fourb\ final state are set for spin-0 and spin-2 resonances 
with values ranging between 24 and 113 fb (at 95\% CL)
for resonance masses in the range between 600 and 3000 \GeV{}.
For nonresonant production, the upper limit is 1.22 pb (at 95\% CL).
The search sensitivity of this analysis exceeds that achieved
at $\rts = 8$ \TeV{} with 19.5~\ifb\ for resonance masses above 1350 \GeV{}.
Furthermore, the search has been extended to cover the mass range between
2000 and 3000 \GeV{}.

\section*{Acknowledgements}

We thank CERN for the very successful operation of the LHC, as well as the
support staff from our institutions without whom ATLAS could not be
operated efficiently.

We acknowledge the support of ANPCyT, Argentina; YerPhI, Armenia; ARC, Australia; BMWFW and FWF, Austria; ANAS, Azerbaijan; SSTC, Belarus; CNPq and FAPESP, Brazil; NSERC, NRC and CFI, Canada; CERN; CONICYT, Chile; CAS, MOST and NSFC, China; COLCIENCIAS, Colombia; MSMT CR, MPO CR and VSC CR, Czech Republic; DNRF and DNSRC, Denmark; IN2P3-CNRS, CEA-DSM/IRFU, France; GNSF, Georgia; BMBF, HGF, and MPG, Germany; GSRT, Greece; RGC, Hong Kong SAR, China; ISF, I-CORE and Benoziyo Center, Israel; INFN, Italy; MEXT and JSPS, Japan; CNRST, Morocco; FOM and NWO, Netherlands; RCN, Norway; MNiSW and NCN, Poland; FCT, Portugal; MNE/IFA, Romania; MES of Russia and NRC KI, Russian Federation; JINR; MESTD, Serbia; MSSR, Slovakia; ARRS and MIZ\v{S}, Slovenia; DST/NRF, South Africa; MINECO, Spain; SRC and Wallenberg Foundation, Sweden; SERI, SNSF and Cantons of Bern and Geneva, Switzerland; MOST, Taiwan; TAEK, Turkey; STFC, United Kingdom; DOE and NSF, United States of America. In addition, individual groups and members have received support from BCKDF, the Canada Council, CANARIE, CRC, Compute Canada, FQRNT, and the Ontario Innovation Trust, Canada; EPLANET, ERC, FP7, Horizon 2020 and Marie Sk{\l}odowska-Curie Actions, European Union; Investissements d'Avenir Labex and Idex, ANR, R{\'e}gion Auvergne and Fondation Partager le Savoir, France; DFG and AvH Foundation, Germany; Herakleitos, Thales and Aristeia programmes co-financed by EU-ESF and the Greek NSRF; BSF, GIF and Minerva, Israel; BRF, Norway; Generalitat de Catalunya, Generalitat Valenciana, Spain; the Royal Society and Leverhulme Trust, United Kingdom.

The crucial computing support from all WLCG partners is acknowledged gratefully, in particular from CERN, the ATLAS Tier-1 facilities at TRIUMF (Canada), NDGF (Denmark, Norway, Sweden), CC-IN2P3 (France), KIT/GridKA (Germany), INFN-CNAF (Italy), NL-T1 (Netherlands), PIC (Spain), ASGC (Taiwan), RAL (UK) and BNL (USA), the Tier-2 facilities worldwide and large non-WLCG resource providers. Major contributors of computing resources are listed in Ref.~\cite{ATL-GEN-PUB-2016-002}.


\clearpage
\pagebreak

\printbibliography



\newpage
\begin{flushleft}
{\Large The ATLAS Collaboration}

\bigskip

M.~Aaboud$^\textrm{\scriptsize 136d}$,
G.~Aad$^\textrm{\scriptsize 87}$,
B.~Abbott$^\textrm{\scriptsize 114}$,
J.~Abdallah$^\textrm{\scriptsize 65}$,
O.~Abdinov$^\textrm{\scriptsize 12}$,
B.~Abeloos$^\textrm{\scriptsize 118}$,
R.~Aben$^\textrm{\scriptsize 108}$,
O.S.~AbouZeid$^\textrm{\scriptsize 138}$,
N.L.~Abraham$^\textrm{\scriptsize 150}$,
H.~Abramowicz$^\textrm{\scriptsize 154}$
% ...
\bigskip
\\
% ...
$^{12}$ Institute of Physics, Azerbaijan Academy of Sciences, Baku, Azerbaijan\\
% ...
$^{65}$ University of Iowa, Iowa City IA, United States of America\\
% ...
$^{87}$ CPPM, Aix-Marseille Universit{\'e} and CNRS/IN2P3, Marseille, France\\
% ...
$^{108}$ Nikhef National Institute for Subatomic Physics and University of Amsterdam, Amsterdam, Netherlands\\
% ...
$^{114}$ Homer L. Dodge Department of Physics and Astronomy, University of Oklahoma, Norman OK, United States of America\\
% ...
$^{118}$ LAL, Univ. Paris-Sud, CNRS/IN2P3, Universit{\'e} Paris-Saclay, Orsay, France\\
% ...
$^{136}$ $^{(a)}$ Facult{\'e} des Sciences Ain Chock, R{\'e}seau Universitaire de Physique des Hautes Energies - Universit{\'e} Hassan II, Casablanca; $^{(b)}$ Centre National de l'Energie des Sciences Techniques Nucleaires, Rabat; $^{(c)}$ Facult{\'e} des Sciences Semlalia, Universit{\'e} Cadi Ayyad, LPHEA-Marrakech; $^{(d)}$ Facult{\'e} des Sciences, Universit{\'e} Mohamed Premier and LPTPM, Oujda; $^{(e)}$ Facult{\'e} des sciences, Universit{\'e} Mohammed V, Rabat, Morocco\\
% ...
$^{138}$ Santa Cruz Institute for Particle Physics, University of California Santa Cruz, Santa Cruz CA, United States of America\\
% ...
$^{150}$ Department of Physics and Astronomy, University of Sussex, Brighton, United Kingdom\\
% ...
$^{154}$ Raymond and Beverly Sackler School of Physics and Astronomy, Tel Aviv University, Tel Aviv, Israel\\
% ...
$^{a}$ Also at Department of Physics, King's College London, London, United Kingdom\\
$^{b}$ Also at Institute of Physics, Azerbaijan Academy of Sciences, Baku, Azerbaijan\\
$^{c}$ Also at Novosibirsk State University, Novosibirsk, Russia\\
$^{d}$ Also at TRIUMF, Vancouver BC, Canada\\
$^{e}$ Also at Department of Physics {\&} Astronomy, University of Louisville, Louisville, KY, United States of America\\
% ...
\end{flushleft}


\end{document}